\documentclass[aps,prb,superscriptaddress,twocolumn,preprintnumbers,footinbib]{revtex4-1}

\usepackage{amsmath,amssymb,amsfonts,mathrsfs,bm,dsfont}
\usepackage[all]{xy} 
\usepackage{color}
\usepackage{array}
\usepackage{graphicx}
\usepackage[space]{grffile} 

\usepackage{url}
\usepackage{float}
\usepackage{bibentry}
\usepackage{tcolorbox}
\usepackage[scr=boondoxo]{mathalfa}
\usepackage{xcolor}
\usepackage{braket}
\usepackage{enumitem} 
\usepackage{cleveref}
\usepackage{ragged2e}
\usepackage{diagbox} 

\bibpunct{[}{]}{,}{n}{}{}
\allowdisplaybreaks[4] 

\begin{document}

\title{Engineering chiral topological superconductivity in twisted Ising superconductors}
\author{Xiaodong Hu, Ying Ran}
\affiliation{Department of Physics, Boston College, Chestnut Hill, MA 02467}
\date{\today}

\begin{abstract}
Van der Waals materials like NbSe$_2$ or TaS$_2$ have demonstrated Ising superconductivity down to atomically thin layers. Due to the spin-orbit coupling, these superconductors have the in-plane upper critical magnetic field far beyond the Pauli limit. We theoretically demonstrate that, twisted bilayer Ising superconductors separated by a ferromagnetic buffer layer can naturally host chiral topological superconductivity with Chern numbers, which can be realized in heterostructures like $\mathrm{NbSe_2/CrCl_3/NbSe_2}$. Under appropriate experimental conditions the topological superconducting gap can reach $>0.1$ meV, leading to readily observable signatures such as the quantized thermal Hall transport at low temperatures.  
\end{abstract}

\maketitle

\section{Introduction}
	Transition metal dichalcogenides (TMD) are van der Waals (vdW) materials and can be prepared as 2D atomic crystals. They have attracted considerable interest and have demonstrated rich electronic phenomena ranging from charge-density wave order, superconducitivity, exciton formation to the optical control of the valley degrees of freedom \cite{lian2018unveiling,he2020valley,jones2013optical,rivera2016valley,li2021optical}. Strikingly, when prepared in few-layer forms, the so-called Ising superconductors gated-2H-MoS$_2$, 2H-NbSe$_2$, 2H-TaS$_2$, 2H-NbS$_2$ and 2H-TaSe$_2$ have anomalously large in-plane upper critical field, several times beyond the Pauli limit \cite{costanzo2016gate,yan2019thickness,Cho2021,de2018tuning,li2017superconducting,lian2019coexistence}. The physical reason for this behavior can be attributed to strong Ising spin-orbit coupling in these materials, which pins the electrons' spin along the $z$-direction and is much less susceptible to an in-plane magnetic field.  

	Due to their 2D nature, TMD allow the fabrication of various vdW heterostructures with flexible tunability and interplay between electronic structures, superconductivity and magnetism. For instance, the ferromagntic proximity effect in monolayer TMD has been well characterized in heterostructures such as WSe$_2$/CrI$_3$ \cite{cai2019atomically,aivazian2015magnetic,mcguire2017magnetic} via optical probes. More recently, ferromagnetic Josephson junction NbSe$_2$/Cr$_2$Ge$_2$Te$_6$/NbSe$_2$ have been fabricated and investigated, and unconventional Josephson phase is reported \cite{ai2021van,idzuchi2020van}.

	On the other hand, a new paradigm in the engineering of quantum phases of matter has been recently developed based on moir\'{e} patterns \cite{cao2018correlated,cao2018unconventional,bistritzer2011moire} introduced by stacking 2D crystals with twisting angles. Motivated by the discovery of correlated insulators and superconductivity in twisted bilayer graphene, the idea of moir\'{e} engineering has been extended into other materials including TMD \cite{li2021imaging,devakul2021magic,angeli2021gamma,wu2019topological}. 

	Interestingly, recent theories pointed out that twisted bilayer cuprate (Bi2212) may realize chiral topological superconductivity with nonzero Chern numbers \cite{can2021high,song2021doping,zhao2021emergent} --- a novel state of matter that yet to be experimentally confirmed, which has triggered further experimental and theoretical investigations \cite{volkov2020magic,volkov2021josephson}. The crucial mechanism of realizing nontrivial topology here is based on the nodal superconductivity, due to sign changes of the pairing order parameter around the Fermi surface. Intuitively, nodal superconductors are naturally located on the boundary between trivial and nontrivial band topology, and engineering topological phases via perturbations becomes possible. 

	In order to fabricate moir\'{e} structures, vdW 2D crystals are highly desirable. Practically, however, Bi2212 may be the only known nodal superconductor among vdW 2D crystals. This puts forward a serious constraint on the moir\'{e}-engineering of topological superconductivity. For example, the TMD superconductors are known to be nodeless and s-wave superconductors, despite theoretical discussions on the role played by magnetic fluctuations in the pairing mechanism \cite{shaffer2020crystalline}. It would be unfortunate if they cannot be included in the moir\'{e}-engineering of topological superconductivity, especially considering their fabricational flexibililty and tunability.

	Motivated by these experimental and theoretical efforts, we ask the following question: is it possible to moir\'{e}-engineer topological superconductivity in TMD superconductors? The answer is positive. We find that twisted Ising superconductors like 2H-NbSe$_2$ and 2H-TaS$_2$ in the presence of an proximity-induced in-plane Zeeman field (beyond the Pauli limit) and out-of-plane supercurrent can host chiral topological superconductivity with Chern number 12 or 6. The topological phases are found to be robust and occupies physically realizable parameter regimes. Under proper experimental conditions the topological pairing gap is $>0.1$meV.

	Not surprisingly, the mechanism underlying our proposal still involves pairing-gap-closing nodes, required by a trivial-to-topological phase transition. Here the pairing nodes are induced by the in-plane Zeeman field (beyond the Pauli limit) for Ising superconductors with Fermi pockets around the $\Gamma$-point in the momentum space. These Zeeman-field induced nodes were firstly pointed out theoretically by He et.al. \cite{he2018magnetic}, and experimental evidences for such nodal states in the presence of an external magnetic field have been reported \cite{xi2016ising,xing2017ising,de2018tuning,navarro2016enhanced}. 

	Unlike the proposal based on cuprates \cite{can2021high,song2021doping}, here the topological superconductivity in TMD twisted bilayers is limited to low temperatures $\lesssim 1$K. Nevertheless, it is worth mentioning a few advantages of the present proposal. First, different from strongly-correlated cuprates, the TMD Ising superconductivity has been fairly well-understood as conventional s-wave pairing without strong correlations. Namely the low energy physics of TMD are well under control in terms of theoretical modelling. Second, detecting the chiral majorana edge modes is the smoking-gun experiment to identify chiral topological superconductivity. In the present proposal, these edge modes are sharply located at the edge of the ferromagnetic buffer layer, and can be detected using either thermal transport or scanning tunnelling microscope (STM) (see FIG.\ref{fig: proposals} for an illustration). In the proposal based on cuprates, these edge modes will hybridize with the nodal superconductivity due to the irregular shape of the atomically thin flakes, and could be challenging to locate in real space.

\section{Main results}
\begin{figure*} 
		\centering
		\includegraphics[width=1.0\textwidth]{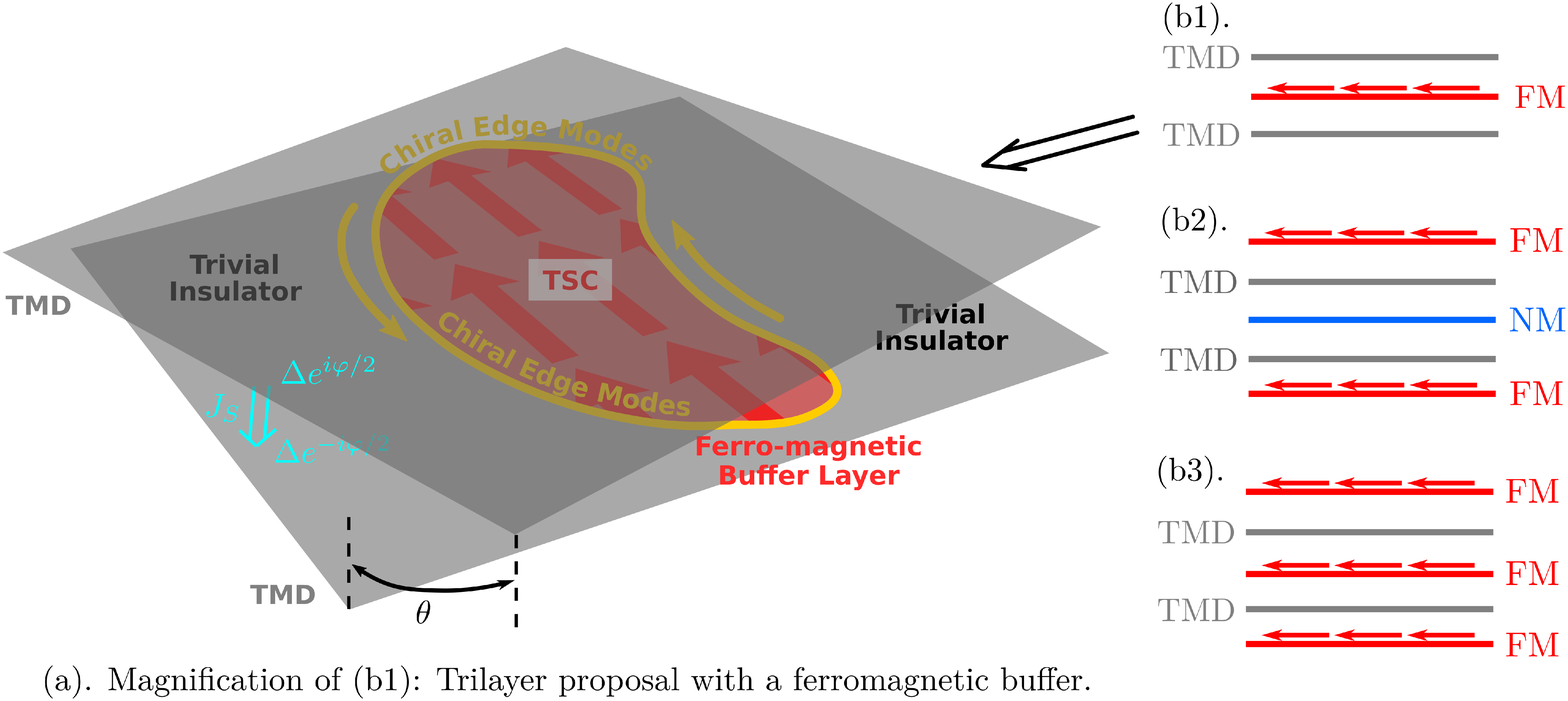}
		\caption{{\bf Setups}: Proposals of twisted $\beta$-bilayer TMD heterostructures with an insulating middle buffer layer. (a,b1): The vertical cyan arrow represents the super-current $J_S$ tuning the Josephson phase between top and bottom layers of TMD, and $\theta$ is a small twisting angle. The middle buffer layer with an \emph{in-plane} ferromagnetic moment (red arrows)  introduces an exchange Zeeman field $H_{\text{exch.}}$ for TMD layers through magnetic proximity effect, realizing the chiral topological superconducting phase. Outside the trilayer overlapping region, trivial superconductivity is realized. The gapless chiral edge modes are localized (yellow curves and arrows) at the boundary of the overlapping region. (b2,b3): Two additional proposals in the same spirits allowing more tunabilities for the parameters $t_\perp,t_{s,\perp}$ and $H_{\text{exch.}}$: FM (NM) means a ferromagnetic (nonmagnetic) buffer layer.}
		\label{fig: proposals}
	\end{figure*}

	Unlike twisted bilayer graphene, twisted heterostructures of TMDs have two distinct configurations differ by a 180$^\circ$ relative rotation, which were referred to as $\alpha$ and $\beta$  \cite{xian2019multiflat}. The $\alpha$-bilayer can be viewed as the building block for the bulk 2H-TMD structure, which restores the inversion symmetry. In this paper we instead focus on $\beta$-bilayer structure of Ising superconductors with Fermi pockets around the $\Gamma$-point (e.g. 2H-NbSe$_2$, 2H-TaS$_2$ but not gated 2H-MoS$_2$). In addition, a ferromagnetic insulating buffer layer with an in-plane magnetic moment is placed in the middle of the $\beta$-bilayer. At small twisting angles between the top and bottom TMD monolayers, we show that the system hosts chiral topological superconductivity when an out-of-plane supercurrent is present (see FIG.\ref{fig: proposals} for an illustration).
	
	Experimentally, most ferromagnetic vdW materials have an out-of-plane magnetic anisotropy. Only recently the monolayer CrCl$_3$ has been successfully isolated and confirmed to have an in-plane ferromagnetic order \cite{cai2019atomically,lu2020meron}. On the other hand, theoretical first-principle calculations predicted that monolayer $\mathrm{Cr_2I_3Cl_3}$ \cite{zhang2020electronic}, 2H-$\mathrm{VS_2}$ \cite{fuh2016newtype} and 2H-$\mathrm{VSe_2}$ \cite{zhang2019magnetic,kezilebieke2020electronic} should be insulators with in-plane ferromagnetic order. In addition, because of a weak magnetic anisotropy, the ferromagnetic moment in CrBr$_3$ can be re-oriented to an in-plane direction by a fairly small external magnetic field ($\sim 0.5$T)\cite{aikebaier2022controlling}. These vdW materials may serve as the ferromagnetic buffer layer in the present proposal.

	Apart from the intrinsic electronic structures of the monolayer TMD, the proposed heterostructures are characterized by three parameters: the Zeeman exchange field $H_{\text{exch.}}$ induced by the magnetic proximity effect, the spin-independent interlayer hopping $t_\perp$, and the spin-dependent interlayer hopping $t_{s,\perp}$. These parameters depend on the choice of the buffer layer in the setup proposed in FIG.\ref{fig: proposals}(a). Moreover, one may consider more sophisticated multilayer setups as shown in FIG.\ref{fig: proposals}(b). By choosing different setups, in principle all the three parameters can be tuned individually.

	In the simplest setup in FIG.\ref{fig: proposals}(a), $H_{\text{exch.}}$ and $t_{s,\perp}$ both are originated from the ferromagnetic buffer layer. In general there is no direct relation between them. However, in a mean-field treatment, the second-order perturbation theory gives $t_{s,\perp}=\mu_B H_{\text{exch.}}$ (see Appendix \ref{app:magnetic_buffer_layer_perturbation}). Although there is no available experimental data to quantify $H_{\text{exch.}}$ for the proposed heterostructures, similar heterostructures like WSe$_2$/CrI$_3$, MoSe$_2$/CrBr$_3$ with out-of-plane ferromagnetic layer have been well characterized experimentally and theoretically \cite{zhong2017van,zhong2020layer,seyler2018valley,zollner2019proximity,ciorciaro2020observation,aikebaier2022controlling}, where a proximity-induced Zeeman splitting around $1\sim2$meV is reported for electronic bands near the $K$ point. A similar value of splitting in the proposed heterostructures corresponds to $H_{\text{exch.}}=3\sim6 H_p$. Notice that monolayer NbSe$_2$ has been reported to sustain a in-plane upper critical field $6\sim 8  H_p$ \cite{xi2016ising,xing2017ising,de2018tuning} (and about $9\sim10 H_p$ for TaS$_2$ \cite{navarro2016enhanced,pan2017enhanced}).

	We find that the topological phase is robust at least when a small twisting angle $\theta$ is comparable with $t_\perp/(\hbar v_{\text{Ising}} k_F)$, where $k_F$ is the Fermi wavevector and the velocity $v_{\text{Ising}}$ is characterizing the Ising spin-orbit coupling near the $\Gamma$-$M$ direction (see Eq.(\ref{eqn:1layer})). It is convenient to introduce a dimensionless parameter to capture the ratio of the two quantities:
	\begin{align}
		\xi\equiv \arctan \frac{2t_\perp}{\hbar v_{\text{Ising}} k_F\theta}.\label{eqn:xi}
	\end{align}

	We consider proposed heterostructures involving either TaS$_2$ or NbSe$_2$. Their monolayer electronic structures are obtained using relaxed crystal structure based on first-principle calculations (see Appendix \ref{app:phase_diagram_details} for details). For each TMD material, two cases for the spin-dependent hopping $t_{s,\perp}$ are investigated, corresponding to:
	
	\begin{align}
	    \textbf{case-(i): }&t_{s,\perp}=\mu_BH_{\text{exch.}}\notag\\
	    \textbf{case-(ii): }&t_{s,\perp}=0\label{eqn:cases}
	\end{align}

\begin{figure*}[!htp]
		\centering
		\begin{minipage}{1.0\textwidth}
			\RaggedRight (a). $\mathbf{TaS_2:}$\\[0.5em]
			\begin{minipage}{0.45\textwidth}
				\centering
				\includegraphics[width=\linewidth]{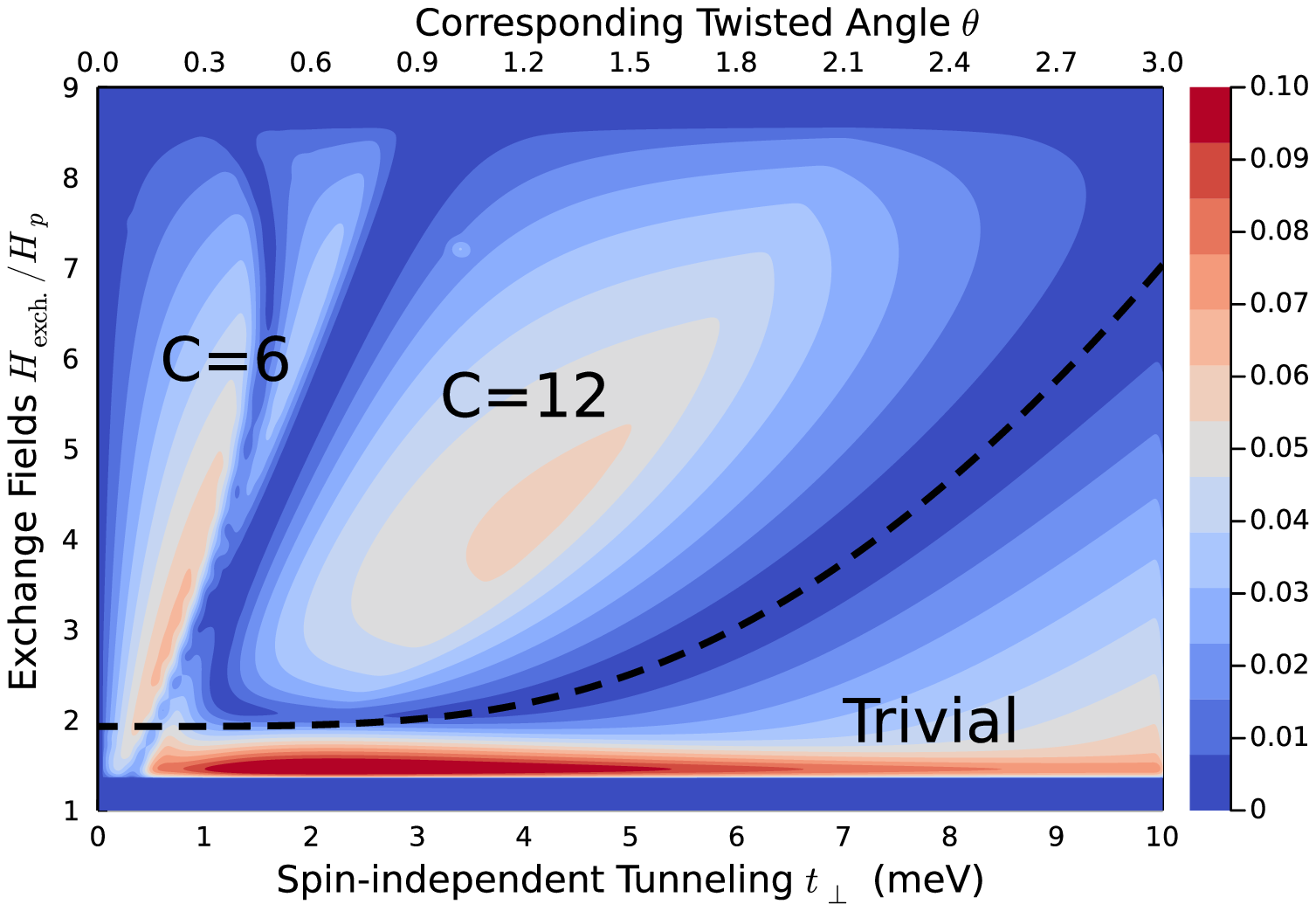}\\
				case (i).
			\end{minipage}
			\hspace{1em}
			\begin{minipage}{0.45\textwidth}
				\centering
				\includegraphics[width=\linewidth]{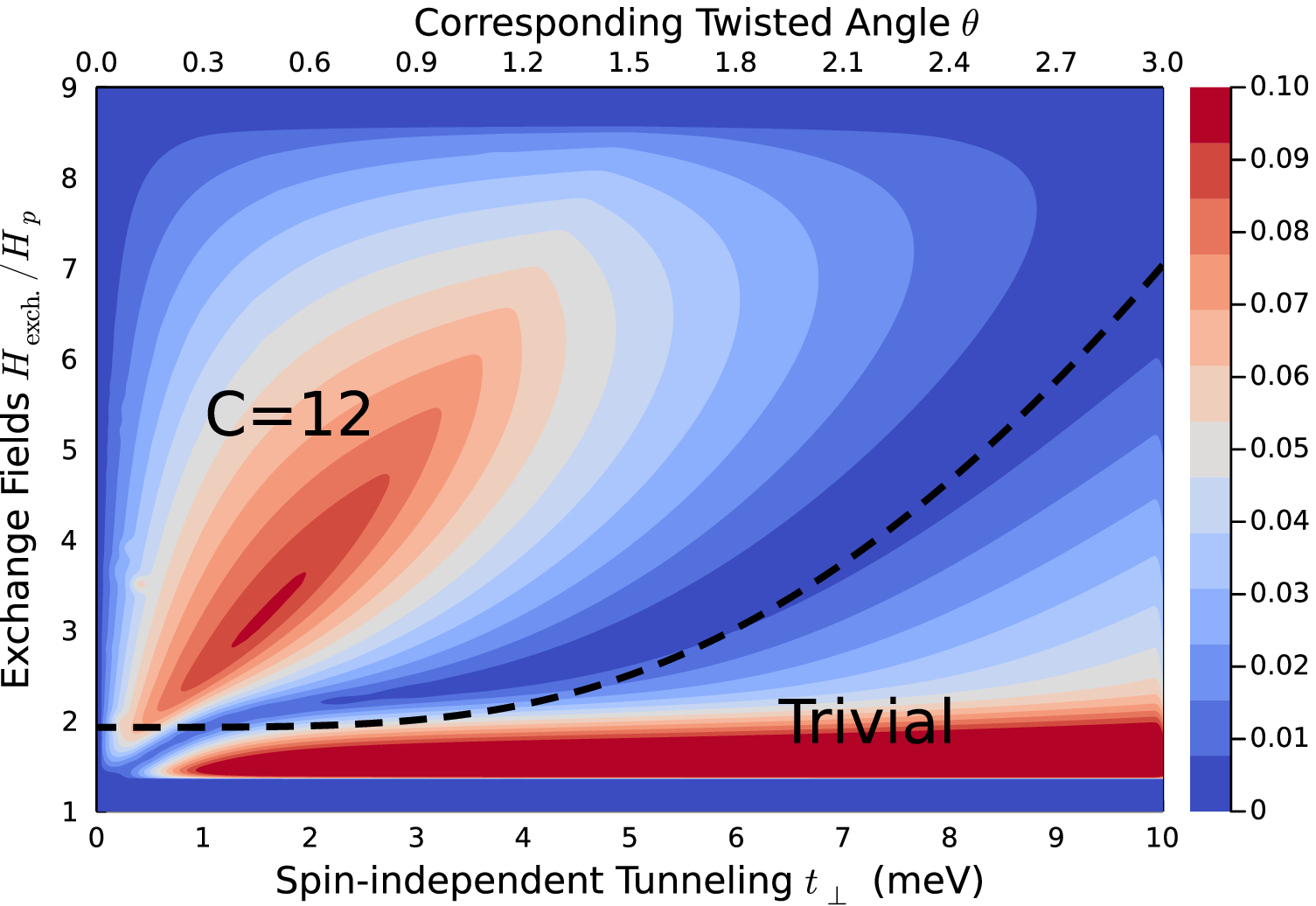}\\
				case (ii).\\
			\end{minipage}
		\end{minipage}
		\\[1em]
		\begin{minipage}{1.0\textwidth}
			\RaggedRight (b). $\mathbf{NbSe2_2:}$\\[0.5em]
			\begin{minipage}{0.45\textwidth}
				\centering
				\includegraphics[width=\linewidth]{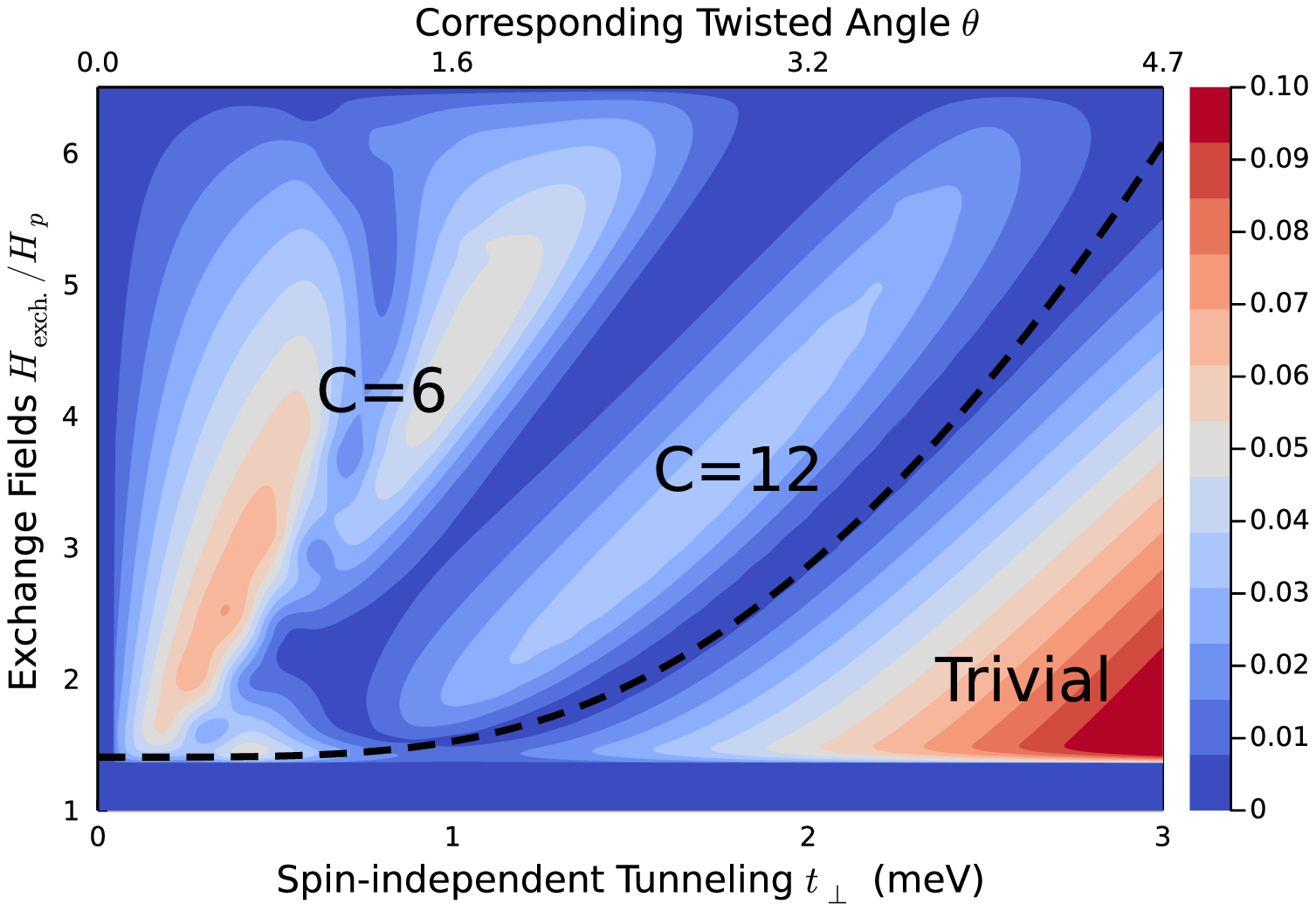}\\
				case (i).
			\end{minipage}
			\hspace{1em}
			\begin{minipage}{0.45\textwidth}
				\centering
				\includegraphics[width=\linewidth]{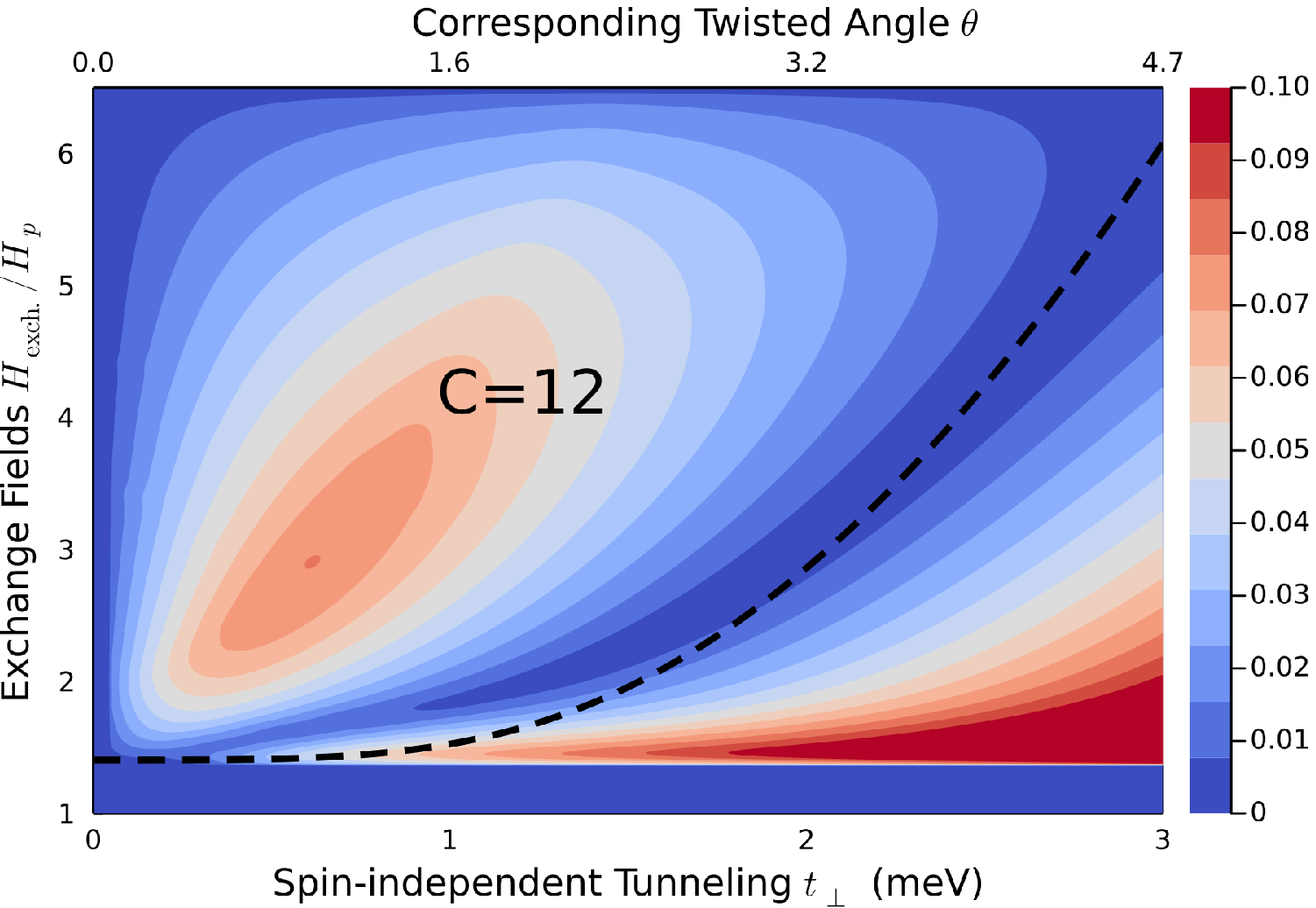}\\
				case (ii).\\
			\end{minipage}
		\end{minipage}
		\caption{{\bf $t_\perp$-$(H/H_p)$ Phase Diagram:} The superconducting gap (gap minimum in the momentum space) and Chern numbers for proposed heterostructures involving $\mathrm{TaS_2}$ (a) and $\mathrm{NbSe_2}$ (b) based on numerical calculations. Here the Josephson phase is fixed to be $\varphi=\pi/2$ and the parameter $\xi$ (see Eq.(\ref{eqn:xi}) for its definition) is fixed to be $\xi^{\mathrm{TaS_2}}=0.9$ and $\xi^{\mathrm{NbSe_2}}=1.1$. $H_p$ in the vertical axis is the Pauli limit field strength. $t_\perp$ and the corresponding twisting angle are displayed on the horizontal axis. For both materials the phase diagrams corresponding to case-(i) and case-(ii) in Eq.(\ref{eqn:cases}) are calculated. The black dashed lines exhibit the predicted topological/trivial phase boundary from analytical perturbative calculations (see Eq.(\ref{Phase boundary})).}
		\label{fig:t-H phase digram}
	\end{figure*}

	We show the global phase diagram by tuning the spin-independent hopping $t_\perp$ and the Zeeman exchange field $H_{\text{exch.}}$ in FIG.\ref{fig:t-H phase digram}, based on numerical calculations. For presentation purpose, we have fixed $\xi^{\mathrm{TaS_2}}=0.9$ and $\xi^{\mathrm{NbSe_2}}=1.1$, and fixed the Josephson phase $\varphi=\pi/2$ corresponding the maximal supercurrent state. Topological superconducting phases are found to exist when $t_\perp\lesssim 9$meV for TaS$_2$ (and $t_\perp\lesssim 2.7$meV for NbSe$_2$). A Chern-number-12 phase is found to occupy a large portion of the parameter space, while a Chern-number-6 phase appears for case-(i) at small values of $t_\perp$. Under appropriate conditions, the gap in the topological phase can reach $0.1$meV. 

	These phase diagrams are also well-understood via analytical calculations. We have plotted the phase boundary (see Eq.(\ref{Phase boundary})) between trivial and topological phases from our perturbative calculations in FIG.\ref{fig:t-H phase digram}, which is in good agreement with the numerical calculations.

\begin{figure*}
			\raggedright(a). $\mathbf{TaS_2:}$\\[1em]
			\begin{minipage}{1.0\textwidth}
				\includegraphics[width=0.29\linewidth]{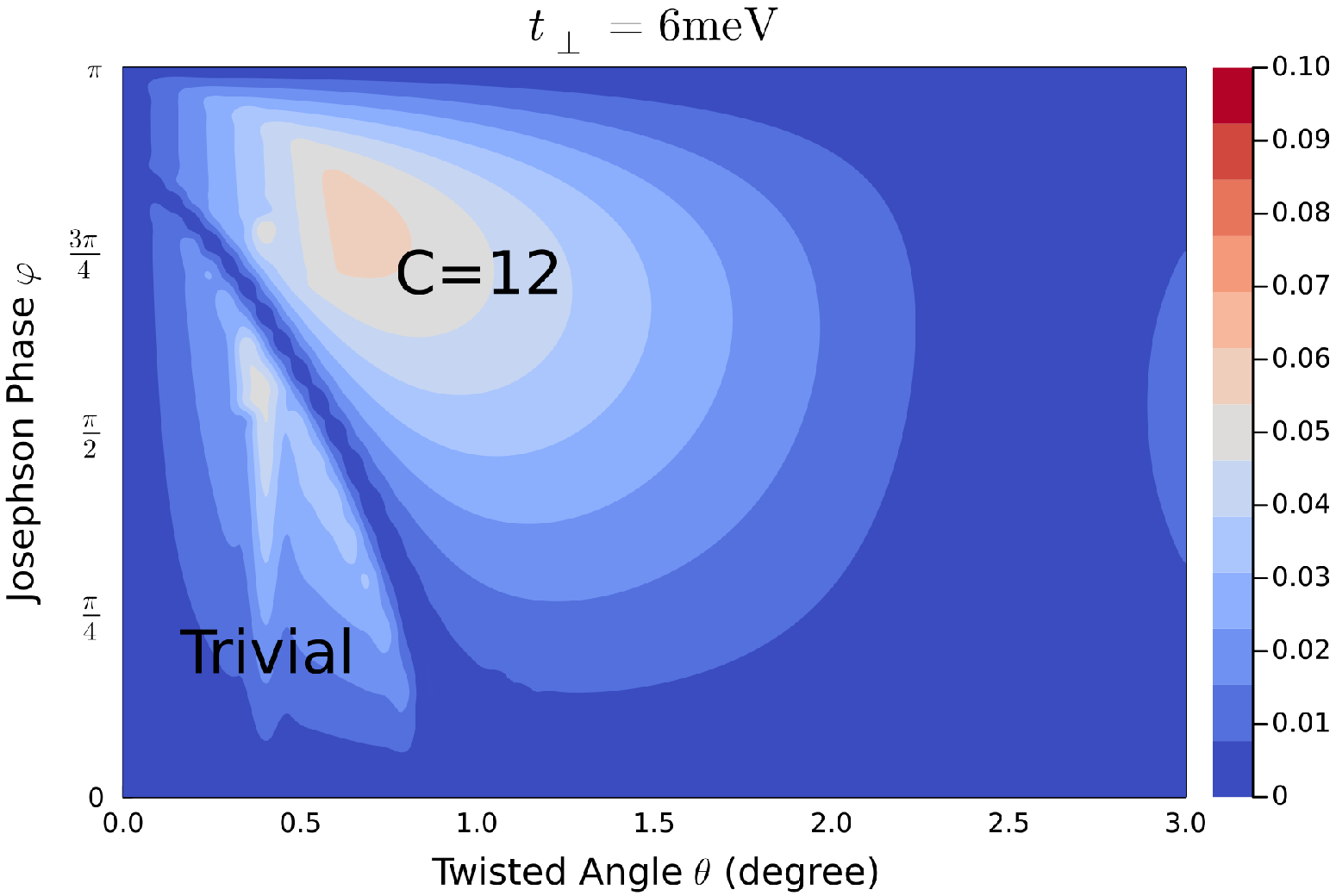}
				\includegraphics[width=0.29\linewidth]{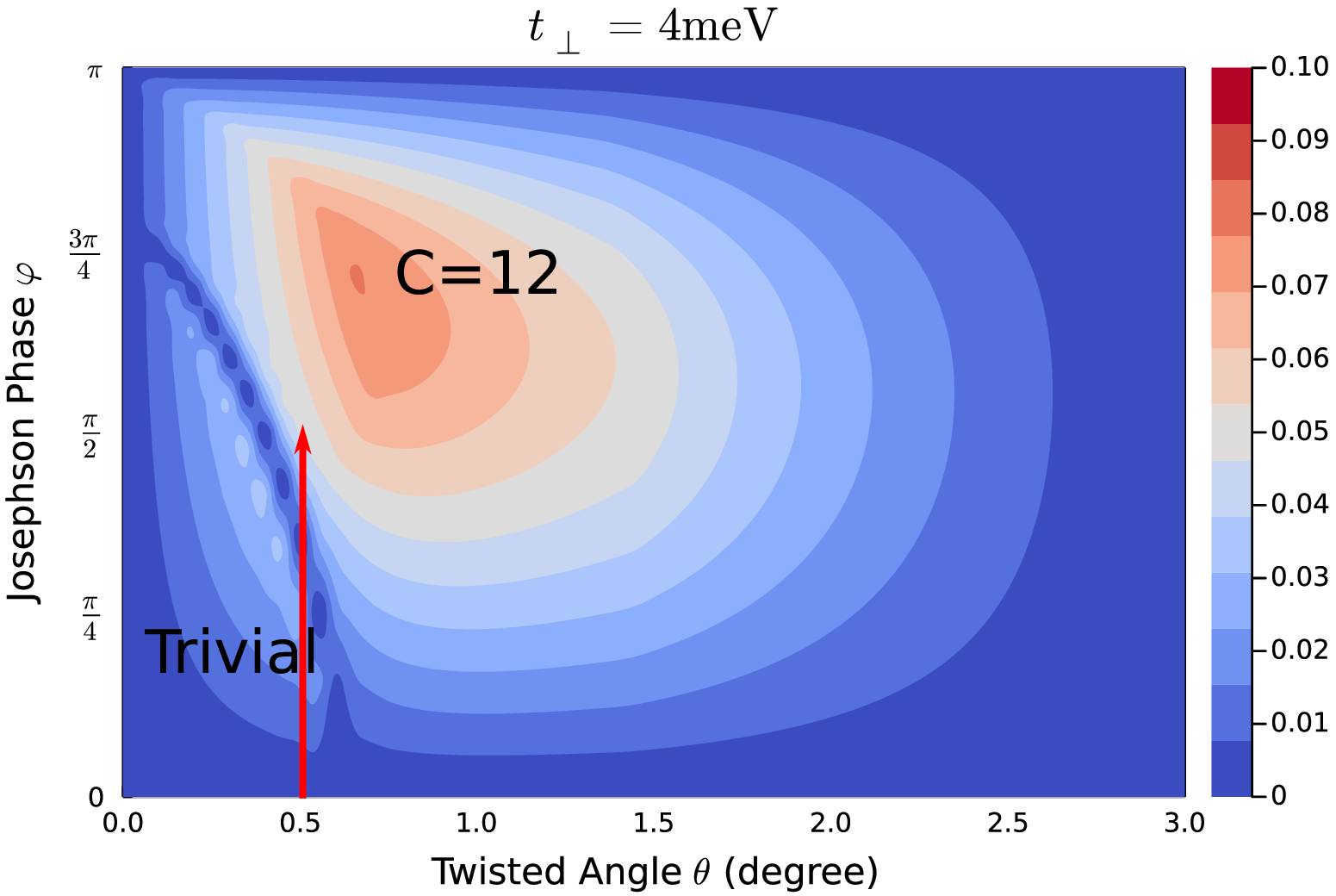}
				\includegraphics[width=0.29\linewidth]{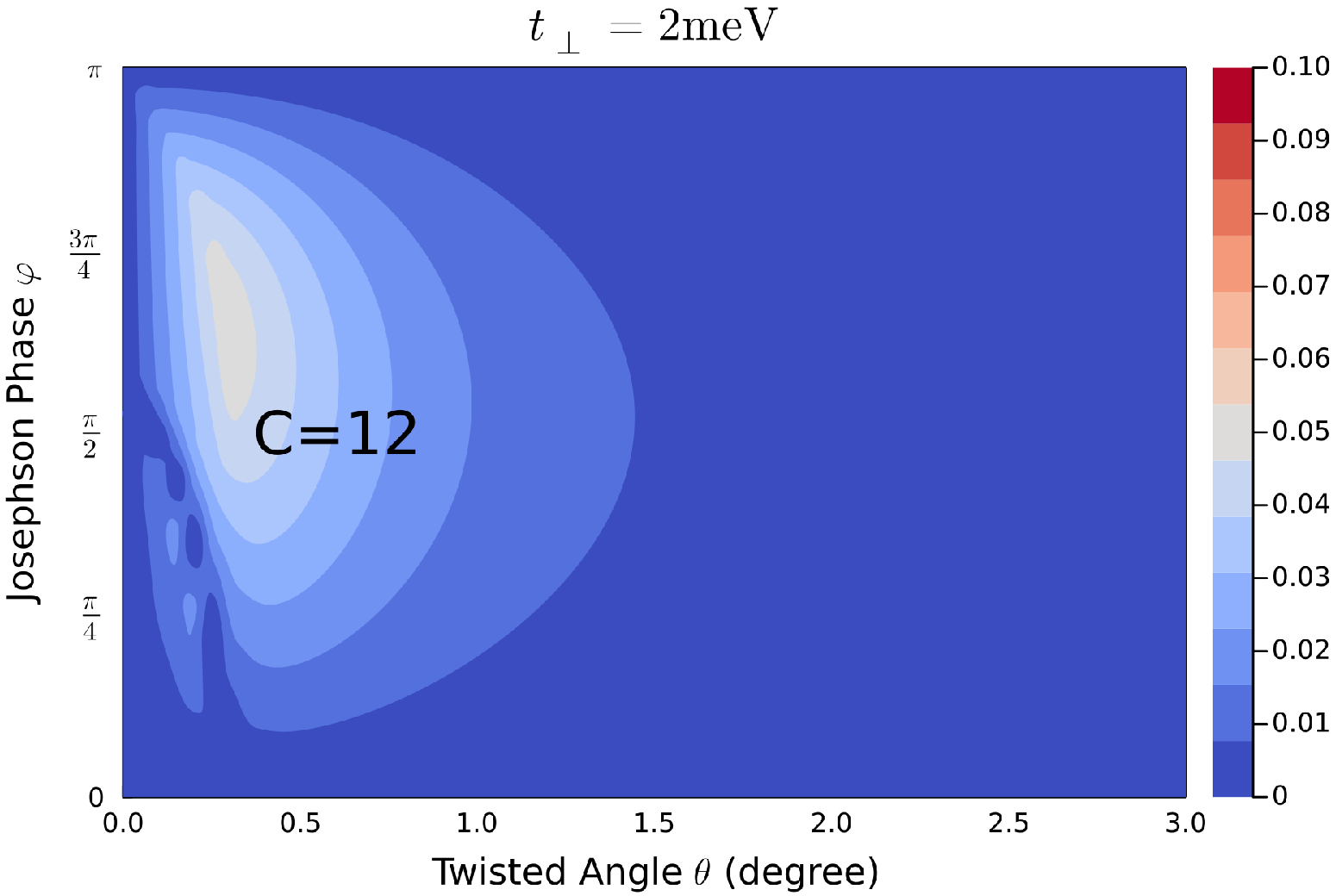}
				\\
				case (i)\\[1em]
				\includegraphics[width=0.29\linewidth]{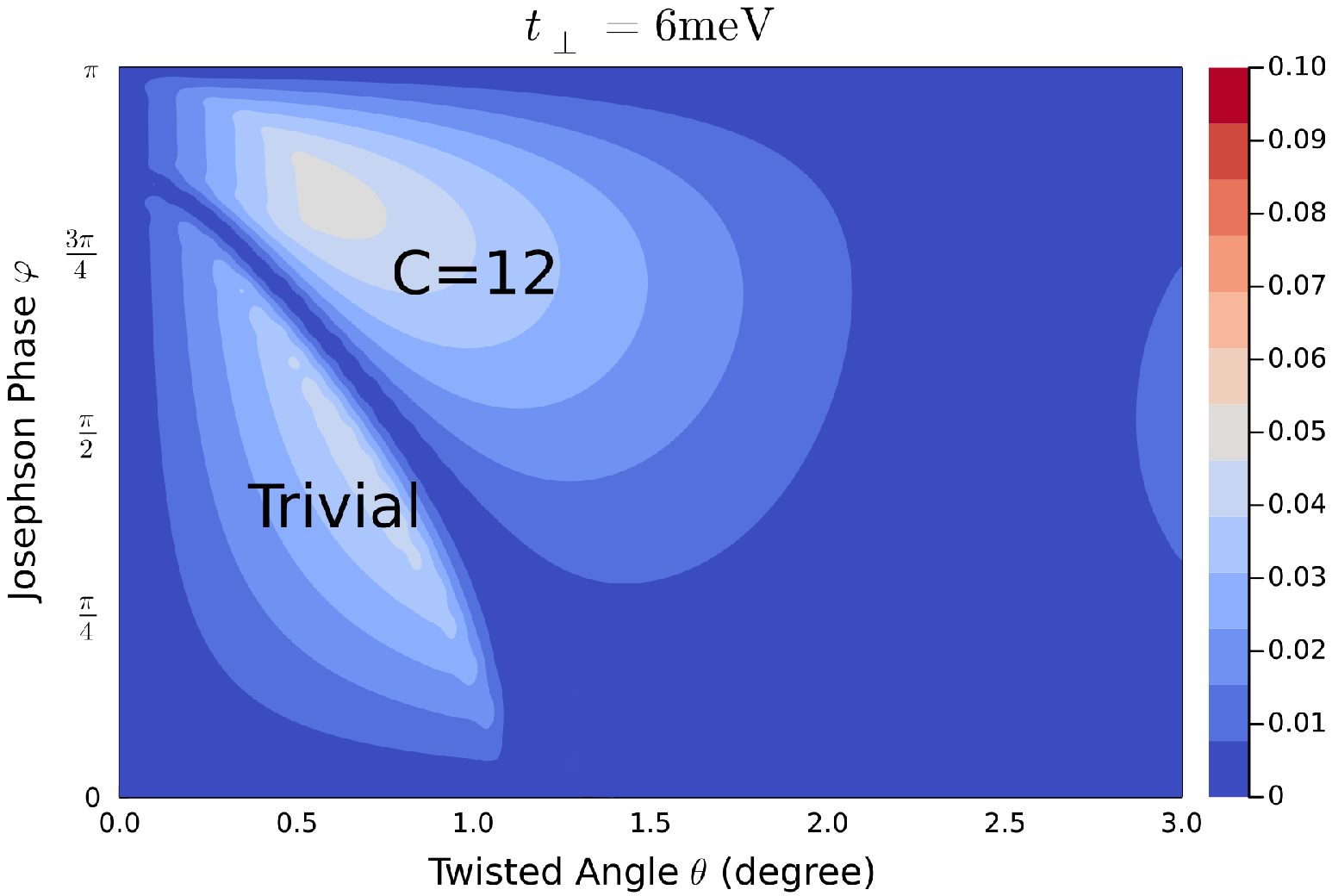}
				\includegraphics[width=0.29\linewidth]{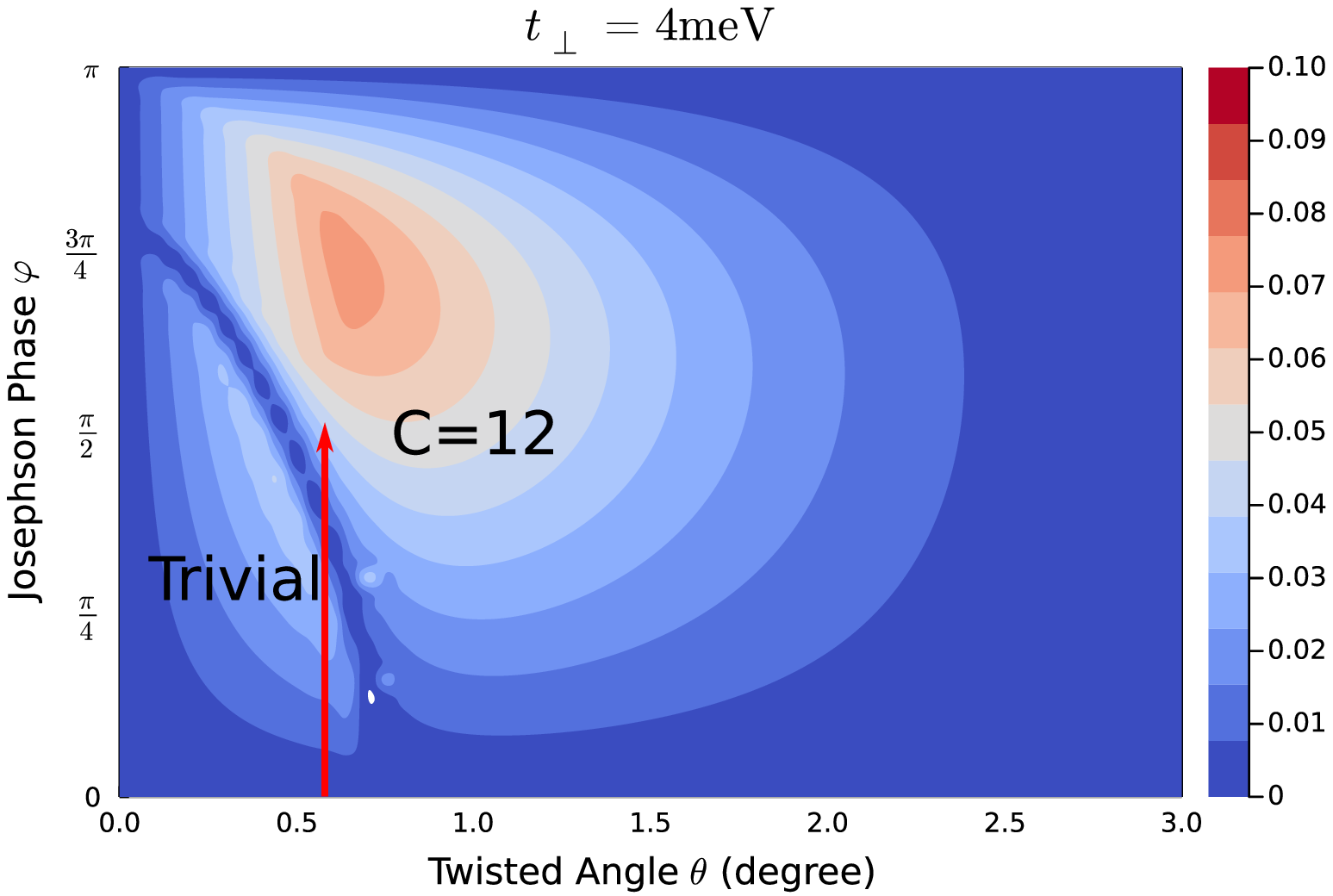}
				\includegraphics[width=0.29\linewidth]{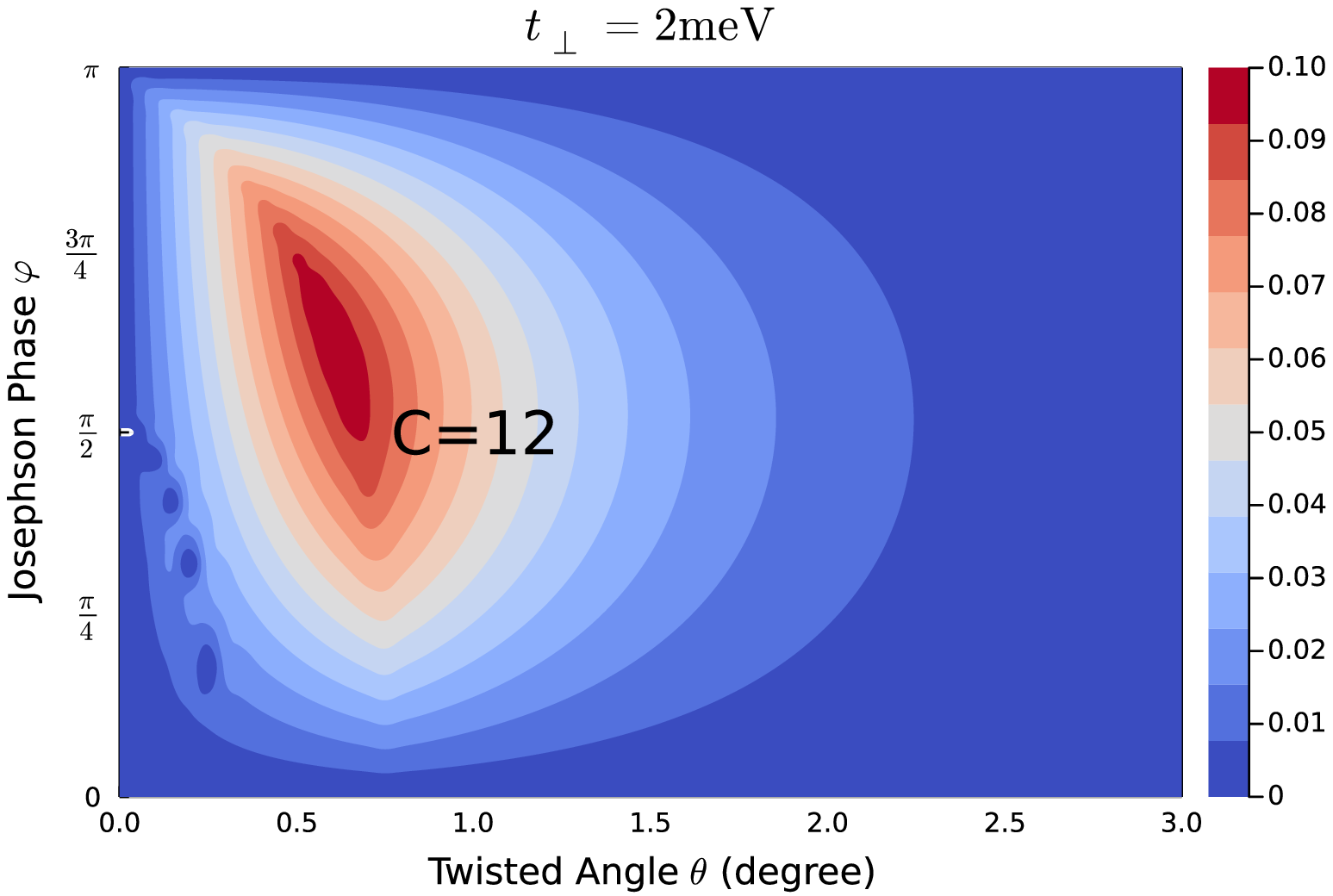}
				\\
				case (ii)
			\end{minipage}
			\\
			\vspace{1.5em}
			\raggedright(b). $\mathbf{NbSe_2:}$\\[1em]
			\begin{minipage}{1.0\textwidth}
				\includegraphics[width=0.29\linewidth]{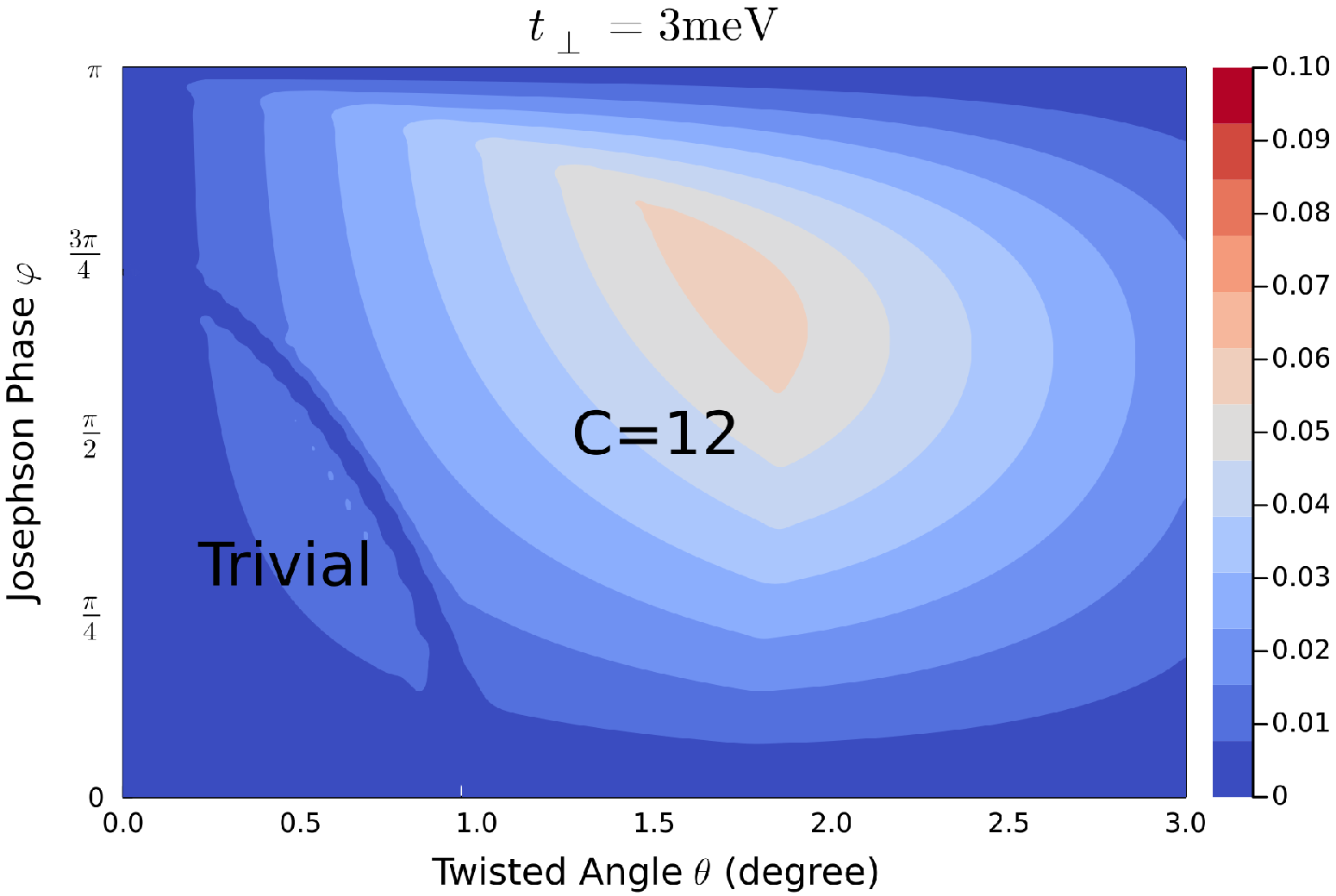}
				\includegraphics[width=0.29\linewidth]{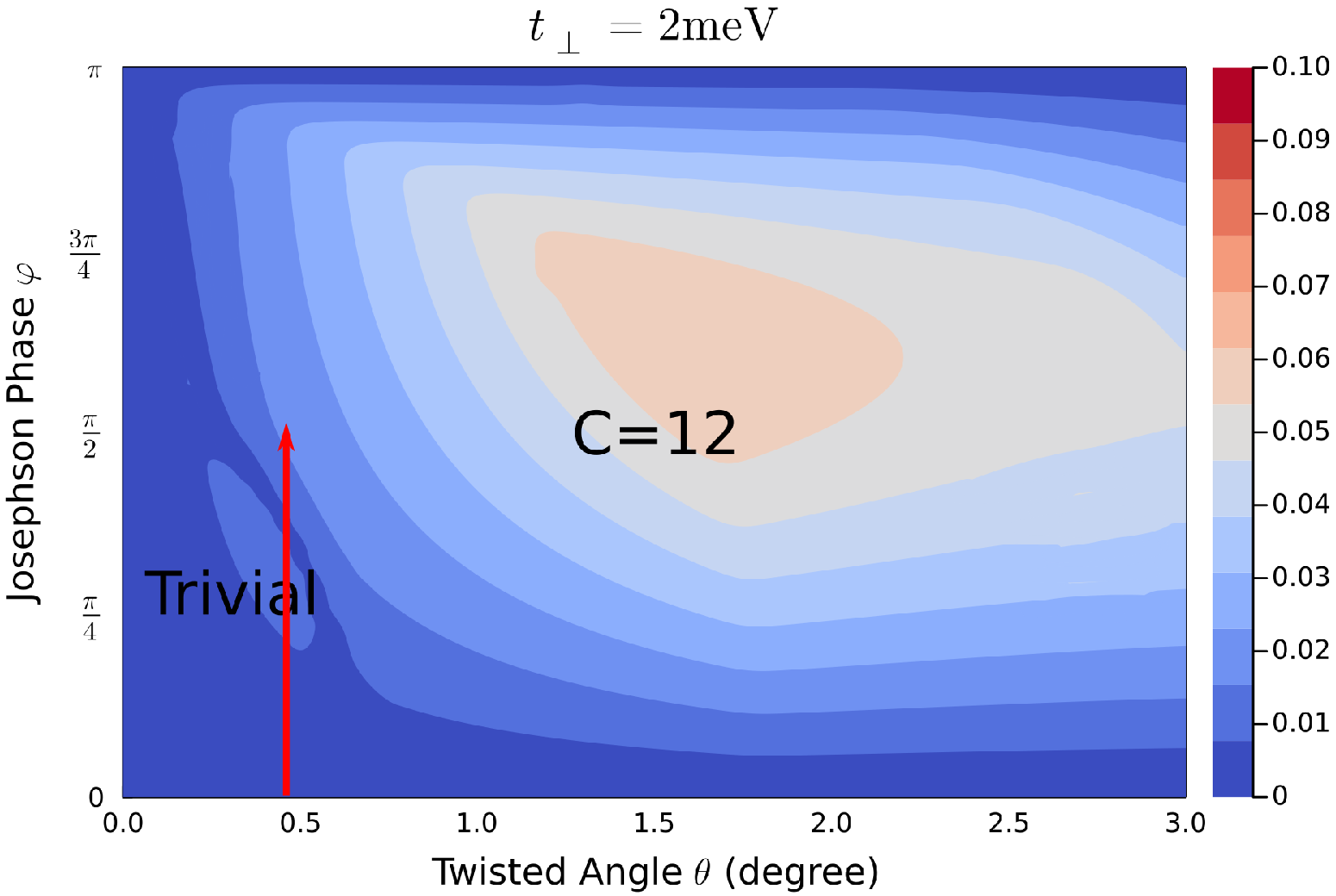}
				\includegraphics[width=0.29\linewidth]{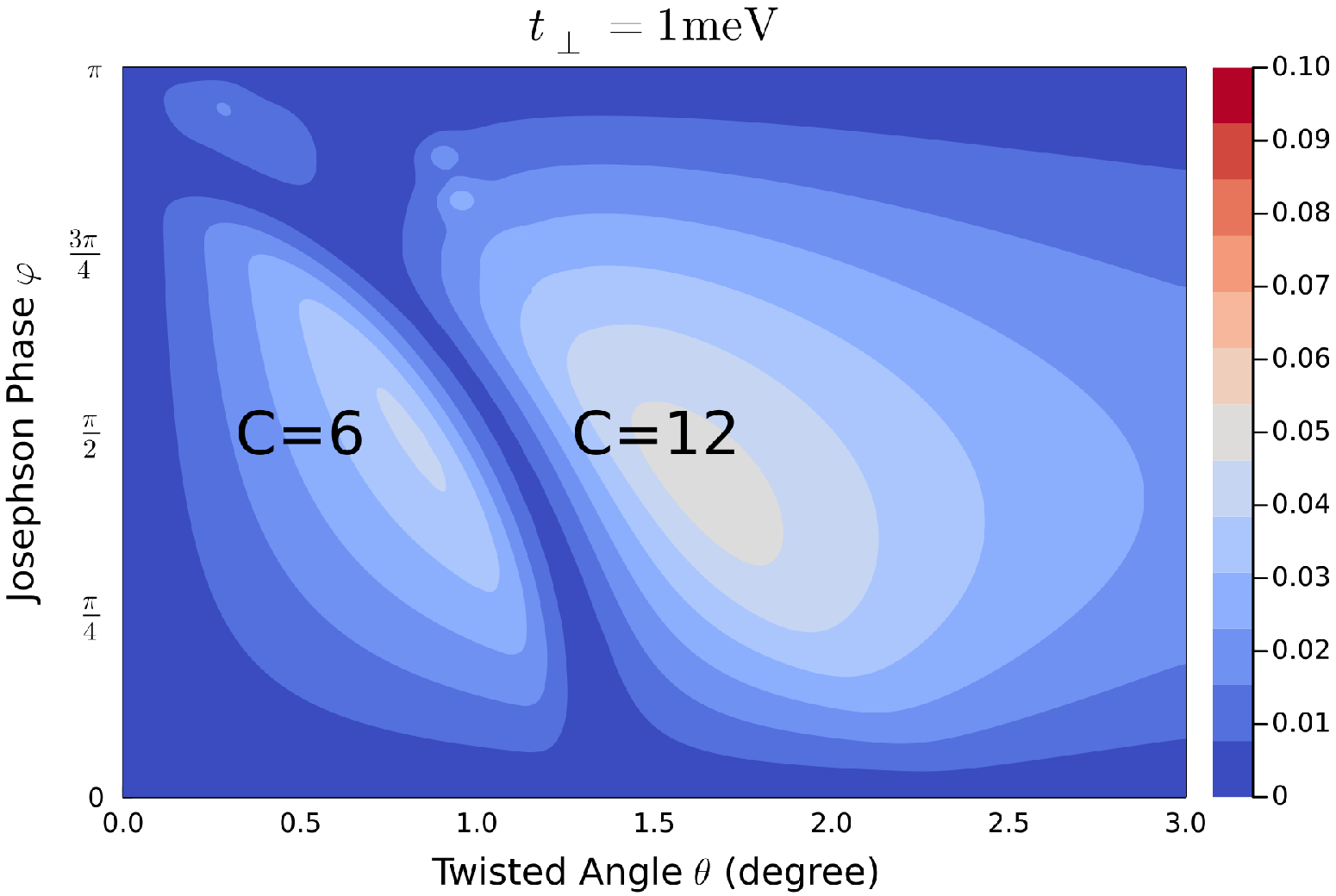}
				\\
				case (i)\\[1em]
				\includegraphics[width=0.29\linewidth]{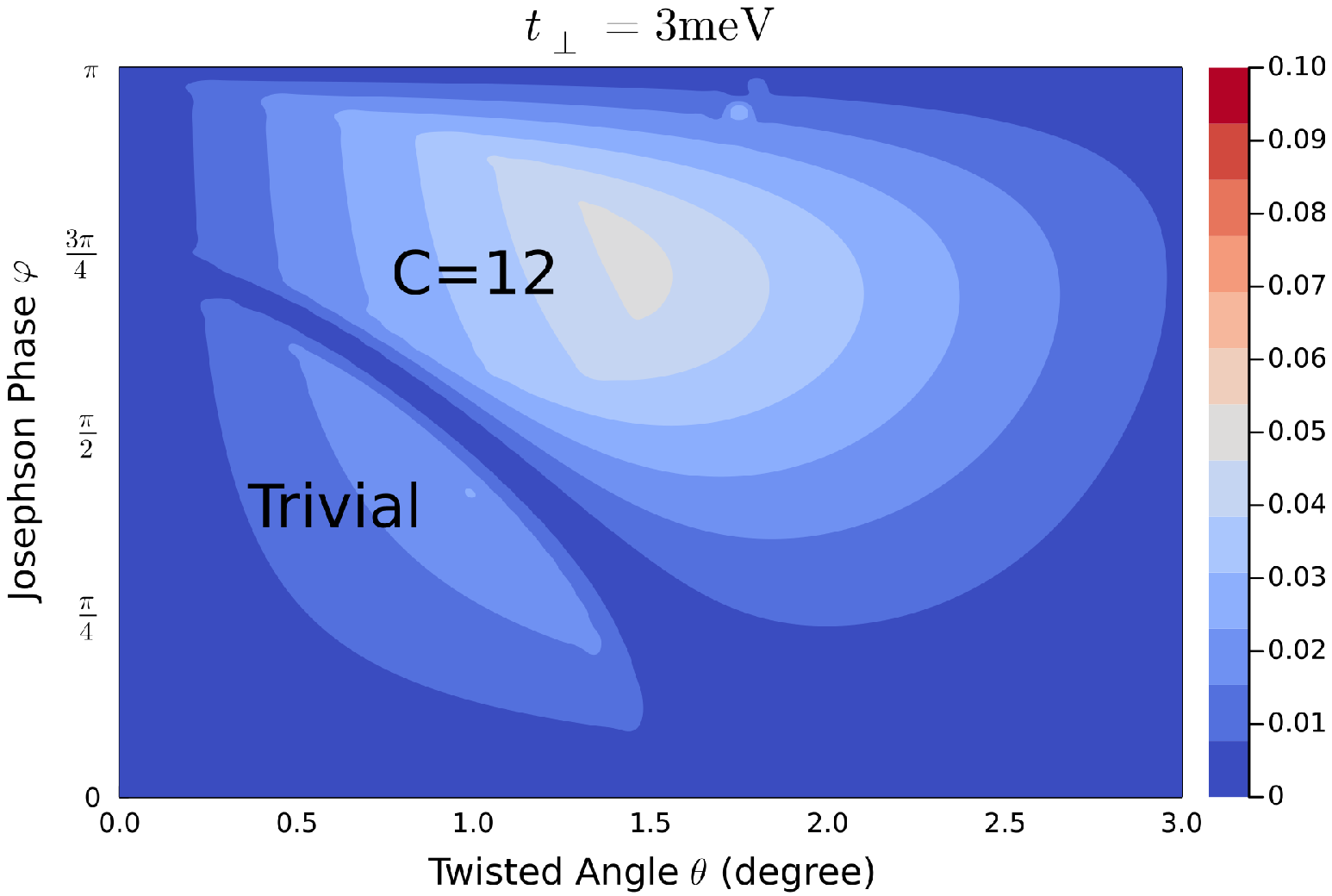}
				\includegraphics[width=0.29\linewidth]{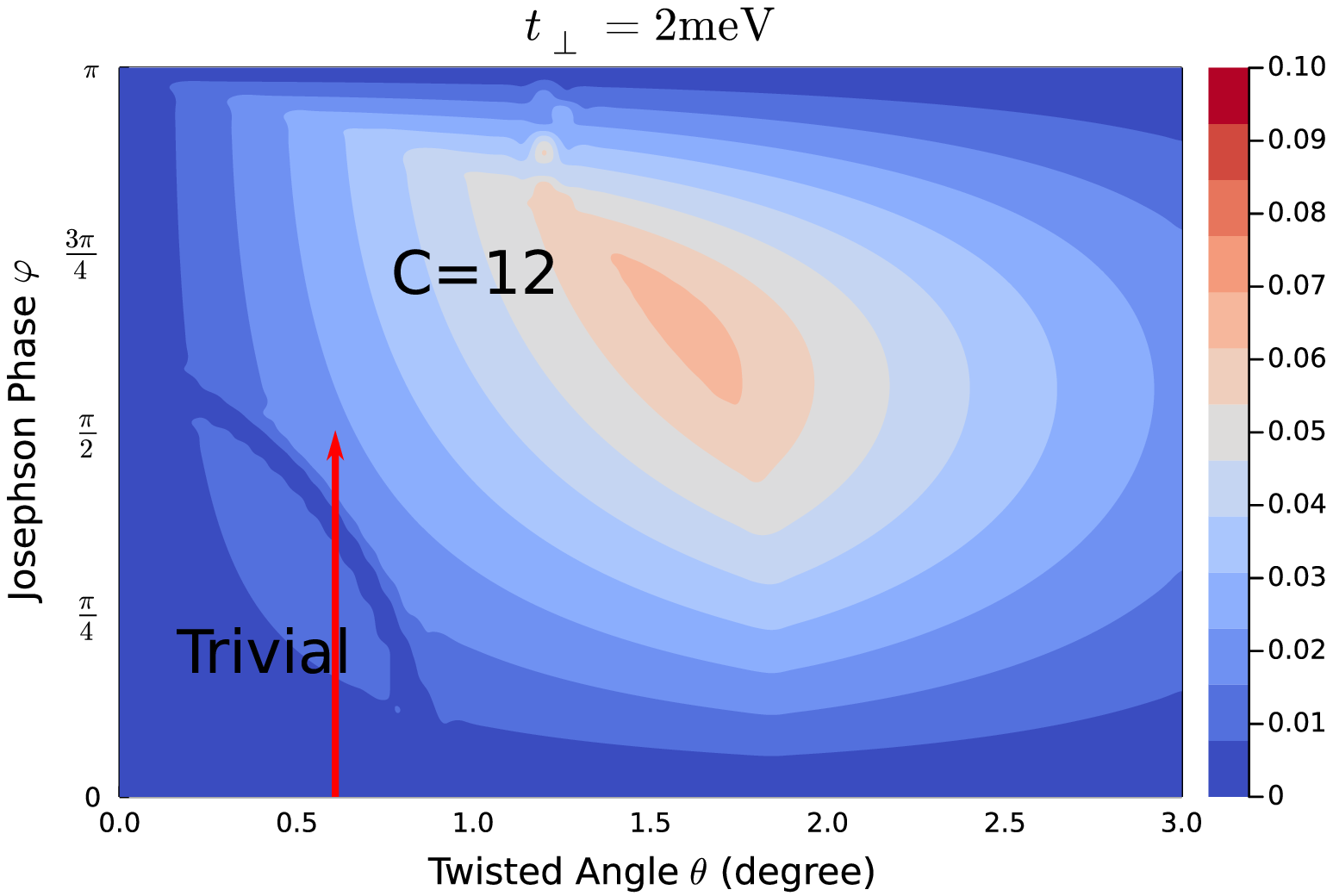}
				\includegraphics[width=0.29\linewidth]{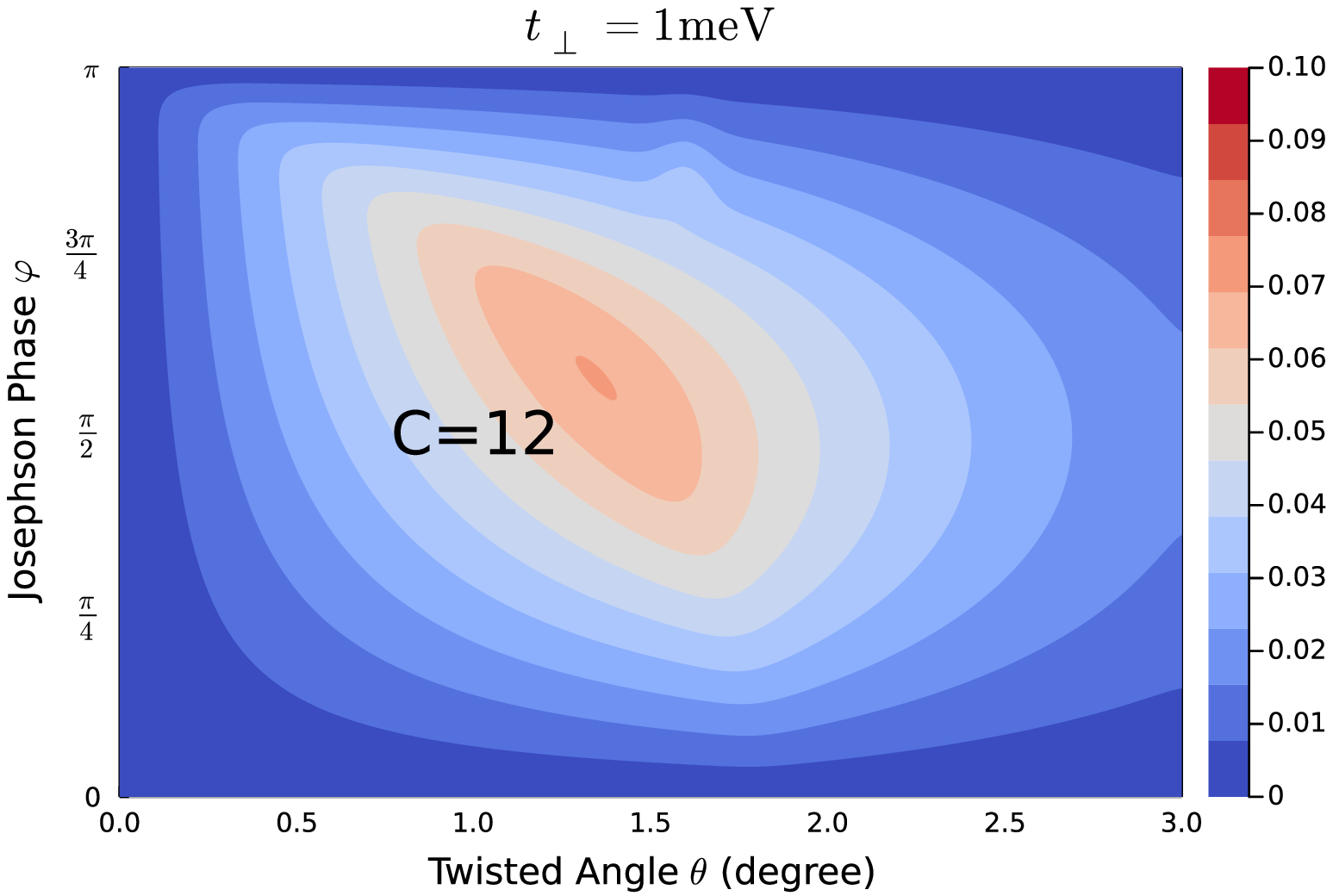}
				\\
				case (ii)
			\end{minipage}
			\caption{{\bf $\theta$-$\varphi$ Phase Diagram:} The superconducting gap (gap mininum in the momentum space) and Chern numbers by varying the twisting angle $\theta$ and the Josephson phase $\varphi$ for proposed heterostructures involving $\mathrm{TaS_2}$ and $\mathrm{NbSe_2}$. Here in each figure we fix the Zeeman field to be four times the Pauli limit $H_{\text{exch.}}=4H_p$ and fix the spin-independent tunneling $t_\perp$. Red arrows are to indicate the trivial-topological phase transitions by tuning $\varphi$ in a single device. Note that the regime $\varphi\leq \pi/2$ corresponds to the stable branch of the Josephson junction.}
			\label{fig:theta-phi phase digram}
	\end{figure*}

	Apart from the global phase diagrams, we also plot the phase diagrams with a few selected values of $t_\perp$. Fixing $H_{\text{exch.}}=4H_p$, we tune the twisting angle $\theta$ and the Josephson phase $\varphi$, and the results are shown in FIG.\ref{fig:theta-phi phase digram}. Consistent with the global phase diagram, the topological phase is realized when $t_\perp$ is not too large. Interestingly, for intermediate values of $t_\perp$, trivial to topological phase transitions are observed by tuning $\varphi$ alone while fixing the twisting angle $\theta$ in an appropriate regime. Such a Josephson-phase-driven topological transition in a single device can lead to unique features to unambiguously identify the topological phase in experiments. For instance, the gapless chiral majorana edge modes should appear only in the topological phase, after the phase transition occurs when $\varphi$ is tuned up.

\section{Model}
	In an in-plane Zeeman field regime $H_{\text{exch.}}>\sqrt{2}H_p$, it was pointed out that the pairing gap around the $\Gamma$-pockets will close and nodal superconductivity is formed \cite{he2018magnetic}, while the pairing gap around the $K$-pockets remain nodeless. These nodes on $\Gamma$-pockets turn out to be the origin of the topological superconductivity in the present proposal. 

	The minimal model for the bands forming the $\Gamma$-pockets (which has dominant $d_{z^2}$ orbital content of the transition metal) is as follows
	\begin{align}
		h(\bm k)=\varepsilon_0(\bm k)+\lambda^{\text{Ising}}_{\mathrm{SO}}(\bm k) \sigma_z.
	\end{align}
	Drastically different from the usual Rashba spin-orbit coupling, here the Ising spin-orbit coupling $\lambda^{\text{Ising}}_{\mathrm{SO}}(\bm k)$ splits bands with fixed $S_z$-spin. In fact, due to the $z\rightarrow -z$ mirror symmetry of monolayer 2H-TMD, the Rashba spin-orbit coupling is forbidden. Time-reversal symmetry dictates that $\lambda^{\text{Ising}}_{\mathrm{SO}}(\bm k)=-\lambda^{\text{Ising}}_{\mathrm{SO}}(-\bm k)$, so $\lambda^{\text{Ising}}_{\mathrm{SO}}(\bm k)$ necessarily has sign changes. Due to the mirror plane containing the $\Gamma$-$M$ axis, the six $\Gamma$-$M$ directions are exact where $\lambda^{\text{Ising}}_{\mathrm{SO}}(\bm k)$ changes sign. 

\begin{figure*}
		\centering
		\raisebox{-0.5\height}{
			\includegraphics[scale=0.75]{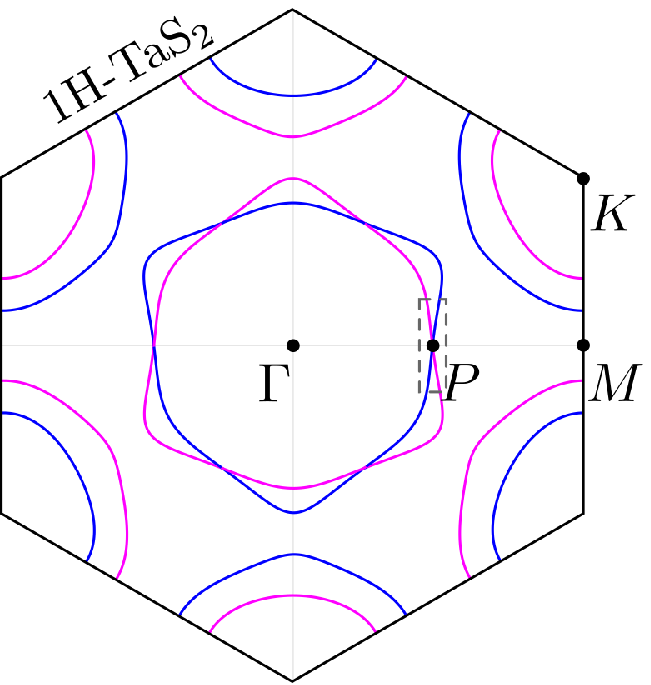}
		}
		\hspace{1.5em}
		\raisebox{-0.5\height}{
			\includegraphics[scale=0.75]{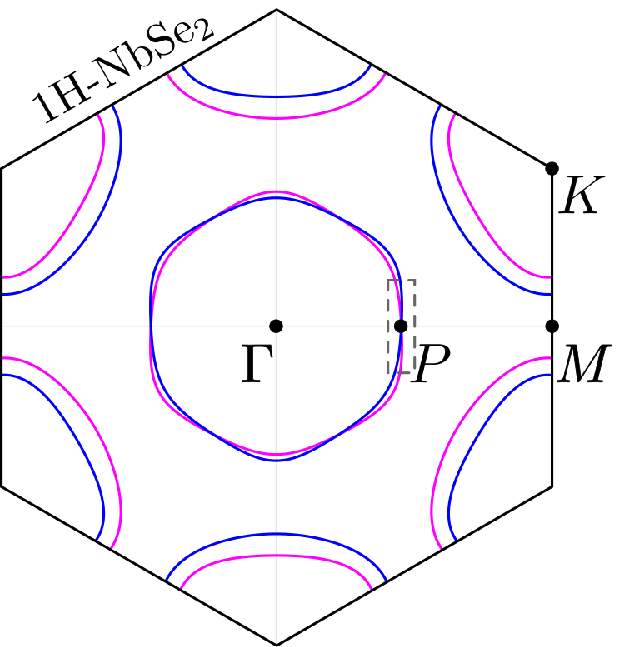}
		}
		\raisebox{-0.5\height}{
			$\xymatrix @+2em { \ar@{=>}[r]^{\text{\small Magnify}}_{\text{\small Dash box}} & }$
		}
		\raisebox{-0.5\height}{
			\includegraphics[scale=0.4]{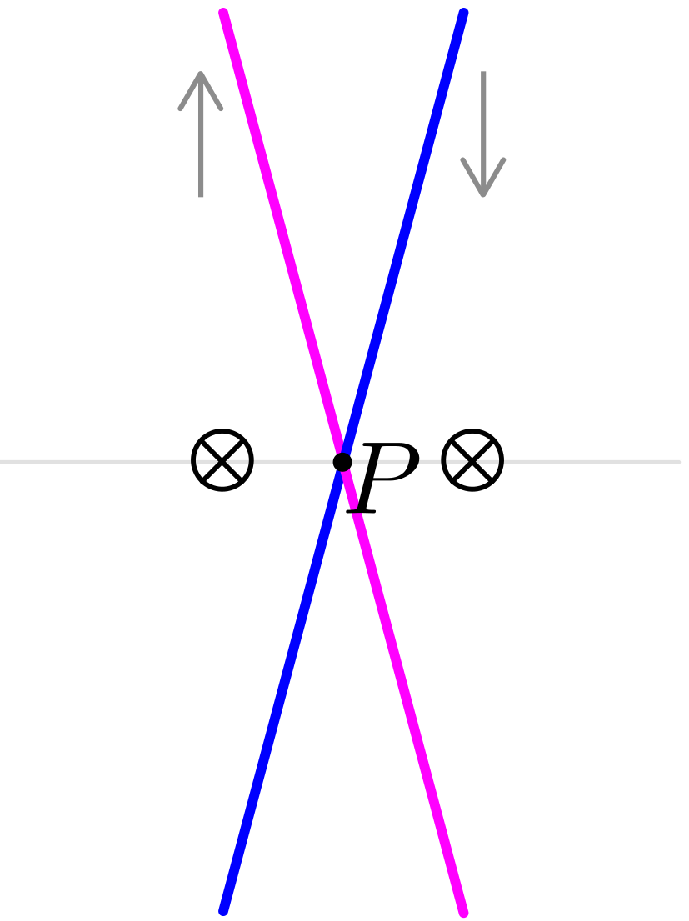}
		}
		\caption{{\bf Monolayer Fermi Surfaces and Band Structures:} $\Gamma$-pocket and $K$-pocket of monolayer $\mathrm{NbSe_2}$ and $\mathrm{TaS_2}$ are shown. Point $P$ is the intersection of $\Gamma$-$M$ line and the the $\Gamma$-pocket, where Ising SOC vanishes and two nodes (cross symbols) emerge when an in-plane Zeeman field $H_{\text{exch.}}$ exceeds the superconducting gap. Magenta and blue colors indicate the different $S_z$-spin components.}
		\label{fig: monolayer FS}
	\end{figure*}

\begin{figure}[!htp]
		\centering
		\begin{minipage}{0.9\linewidth}
			\RaggedRight (a). $\mathbf{TaS_2}$:\\[0.5em]
			\includegraphics[width=\textwidth]{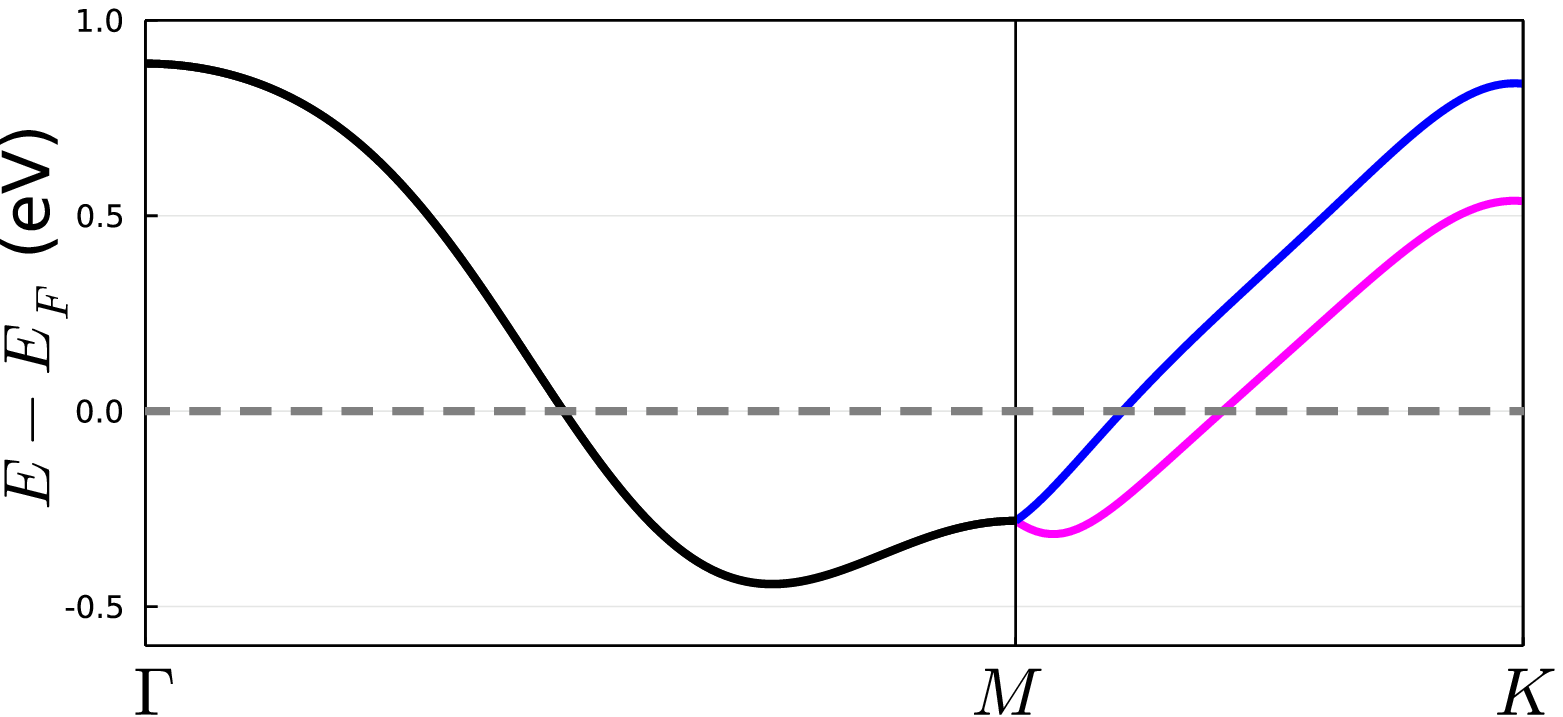}	
		\end{minipage}
		\\[2em]
		\begin{minipage}{0.9\linewidth}
			\RaggedRight (b). $\mathbf{NbSe2_2}$:\\[0.5em]
			\includegraphics[width=\textwidth]{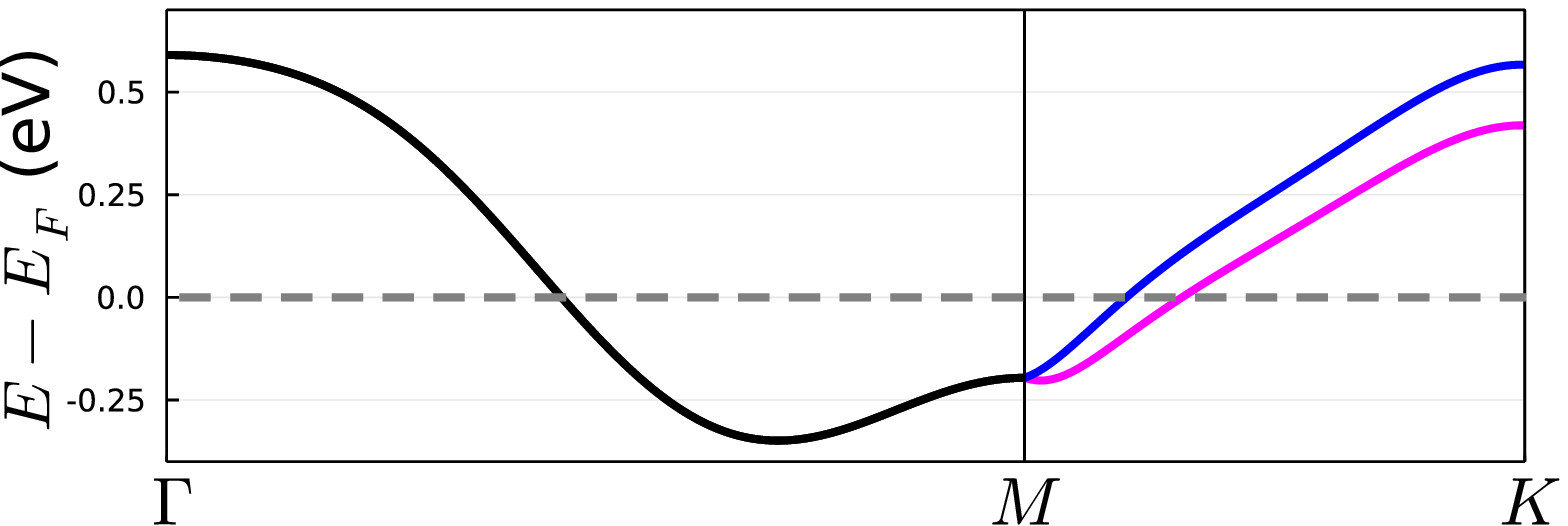}	
		\end{minipage}
		\caption{{\bf Monolayer Band Structures.}}
		\label{fig: band_plot Gamma-M-K}
	\end{figure}

	We therefore begin with a first-order $\bm{k\cdot p}$ model near the intersection point $P$ between the Fermi surface and one $\Gamma$-$M$ direction ($k_x$-direction, see Fig.\ref{fig: monolayer FS} for an illustration):
	\begin{align}
		h^{\text{mono.}}=\hbar v_F k_x + \hbar v_{\text{Ising}} k_y\sigma_z,\label{eqn:1layer}
	\end{align}
	where $\lambda^{\text{Ising}}_{\mathrm{SO}}(\bm k)$ vanishes along the $k_x$-direction, and $v_{\text{Ising}}\equiv \frac{\partial \lambda^{\text{Ising}}_{\mathrm{SO}}(\bm k)}{\hbar \partial k_y}$. We list these parameters for NbSe$_2$ and TaS$_2$ in Table.\ref{table:parameters} based on first principle calculations (Fig. \ref{fig: band_plot Gamma-M-K} plot the band structures, full list of quartic $\bm{k\cdot p}$ expansion parameters is given in Appendix \ref{app:phase_diagram_details}).
	\begin{table}
		\begin{tabular}{|c|c|c|c|c|}
			\hline
			TMD  & $\hbar v_F$(eV$\cdot$\AA) & $\hbar v_{\text{Ising}}$ (eV$\cdot$\AA)  & $k_F$ (\AA$^{-1}$) &$\Delta$ (meV)\\ \hline 
			NbSe$_2$ & -2.22 & 0.12 & 0.48 & 0.46 \\ \hline
			TaS$_2$ & -2.89 & 0.36 & 0.54 & 0.52 \\ \hline
		\end{tabular}
		\caption{The parameters in the $\bm{k\cdot p}$ model.}
		\label{table:parameters}
	\end{table}

	\vspace{1em}
	\textbf{Monolayer:} After introducing an spin-singlet pairing $\Delta$ \footnote{$\Gamma$-$M$ mirror symmetry protects that there is no singlet-triplet mixing at $P$-point.}, in the Nambu basis we obtain the BCS mean-field Hamiltonian:
	\begin{align}\label{eqn:1layer_bcs}
		h^{\text{mono.}}_{\text{BCS}}=&\hbar v_F k_x \tau_z+ \hbar v_{\text{Ising}} k_y\sigma_z+\mu_B H_y \sigma_y+\Delta \sigma_y\tau_y,
	\end{align}
	where Pauli $\tau$-matrices label the particle-hole space, and an Zeeman field $H_y$ along the $y$-direction is introduced (the g-factor is assumed to be 2). Because the system has a spin-rotation symmetry around $S_z$-direction, the choice of the direction of the in-plane magnetic field is not important. Since there are three independent $\Gamma$-$M$-directions, related by $C_3$ rotations, the full electronic structure of NbSe$_2$ (TaS$_2$) has three copies of the effective theory in Eq.(\ref{eqn:1layer_bcs}).
	Eq.(\ref{eqn:1layer_bcs}) can be solved straightforwardly. A pair of Dirac nodes are found to emerge when $\mu_B H_y>\Delta$, located at $\bm k^{\pm}=(\pm \frac{1}{\hbar v_F} \sqrt{(\mu_B H_y)^2-\Delta^2},0)$ (see Fig.\ref{fig: monolayer FS}) for a schematic illustration, consistent with earlier works \cite{he2018magnetic,shaffer2020crystalline}, whose low energy effective theories are:
	\begin{align}
		h^{\pm}_{\text{node}}=\pm\hbar v_F \cos\eta \delta k_x\Sigma_z+\hbar v_{\text{Ising}}\sin\eta \delta k_y\Sigma_x,\label{eqn:1layer_eff}
	\end{align}
	where $\eta=\arcsin\frac{\Delta}{\mu_B H_y}$, $\delta \bm k$ is measured from the nodal points $\bm k^{\pm}$, and $\Sigma$ are Pauli matrices in the low energy space. These nodes are protected by a chiral symmetry $\sigma_x\tau_y$ in model Eq.(\ref{eqn:1layer_bcs}) (corresponding to the combination of the physical time-reversal transformation $i\sigma_yK$, the particle-hole transformation $\tau_x K$ and a $S_z$-$\pi$ rotation $i\sigma_z\tau_z$) sending $h(\bm k)\rightarrow -h(\bm k)$. 
	\vspace{1em}\par
	\textbf{Twisted bilayer:} Next we consider a $\beta$-bilayer (separated by an insulator buffer layer) with a twisting angle $\theta$, which can be viewed as top/bottom monolayer twisted by angle $\pm\theta/2$ respectively. Introducing $k_F$ being the crystal momentum of point $P$, we obtain the following effective theory near $P$:
	\begin{align}
		h^{\text{bilayer}}_{\text{BCS}}=&\hbar v_F k_x\tau_z+\hbar v_{\text{Ising}} k_y\sigma_z+\Delta\cos\frac{\varphi}{2}\sigma_y\tau_y\notag\\
		+&\Delta\sin\frac{\varphi}{2} \sigma_y\tau_x\nu_z-\hbar v_{\text{Ising}}k_F\frac{\theta}{2}\sigma_z\nu_z+\mu_B H_y\sigma_y\notag\\
		+&t_\perp \tau_z\nu_x+t_{s_y,\perp}\sigma_y\nu_x.\label{eqn:bilayer_bcs}
	\end{align}
	Here the Pauli matrices $\nu$ captures the top/bottom layer space. $\Delta>0$ and the pairing amplitude is $\Delta e^{\pm i\varphi/2}$ for the top/bottom layer respectively. The $t_\perp$ and $t_{s_y,\perp}$ terms describes the spin-independent and spin-dependent interlayer hopping processes respectively. Based on the two-center approximation \cite{bistritzer2011moire,devakul2021magic}, and the fact that point $P$ is far away from the Brillouin zone boundary, the interlayer hopping processes have a weak stacking-dependence, and are considered as constants in the present work \cite{volkov2020magic,volkov2021josephson} (see Appendix \ref{app:two-center approximation} for a discussion on the two-center approximation).

	The model Eq.(\ref{eqn:bilayer_bcs}) can be analytically solved in various perturbative regimes. To demonstrate the stability of the topological superconductivity, below we focus on one particular regime, in which $\sqrt{(\hbar v_{\text{Ising}} k_F\theta)^2+4t_\perp^2}\gg \Delta,\mu_B H_y, t_{s_y,\perp}$. The advantage of this regime is that it is always realizable in the proposed heterostructures by tuning the twisting angle $\theta$. In this regime, we find that the topological superconductivity with Chern number 4 (corresponding to Chern number 12 for the whole heterostructure) is always realized in model Eq.(\ref{eqn:bilayer_bcs}) by tuning $\varphi,\theta$ and $\mu_B H_y \gtrsim  \Delta$.

\begin{figure}[!htp]
		\centering
		\includegraphics[scale=0.4]{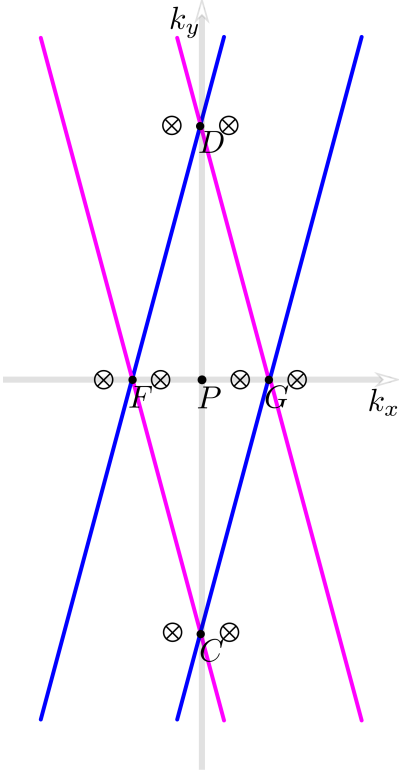}
		\caption{{\bf Schematic Plot of Four Fermi Surface Intersection Points in a Twisted Bilayer Near Point-P}: When $\varphi=0$, each intersection point will give rise to two pairing nodes after $H_{\text{exch.}}$ is tuned up. When $\varphi\neq 0$, these nodes open up energy gaps: The four nodes near points $F$ and $G$ along the $\Gamma$-$M$ line have total Chern number zero. The four nodes near points $C$ and $D$ have total Chern number 4.}
		\label{fig: all eight nodes}
	\end{figure}

	To understand this behavior, we may firstly turn off the pairing $\Delta$ and $\mu_B H_y, t_{s_y,\perp}$  in model Eq.(\ref{eqn:bilayer_bcs}). There are four intersection points between the spin-up and spin-down Fermi surfaces, which we label as $C,D$ and $F,G$ and are located at the $k_y$ and $k_x$ axes respectively (see FIG. \ref{fig: all eight nodes}).
	The low energy effective theory near each point resembles Eq.(\ref{eqn:1layer}) with modified $v_F$ and $v_{\text{Ising}}$.

	After $\Delta,\varphi$ are turned on but $\mu_B H_y=t_{s_y,\perp}=0$, the pairing gap minima near $C,D,F,G$ are found to be the same $\Delta_{C,D,F,G}=\Delta\sqrt{\cos^2\frac{\varphi}{2}+\cos^2\xi\sin^2\frac{\varphi}{2}}$, where $\xi$ is defined in Eq.(\ref{eqn:xi}).
	Because the connection with the $\varphi=0$ limit, the superconducting phase so far must still be topologically trivial even though $\varphi\neq 0$ breaks the time-reversal symmetry.

	When $\mu_B H_y,t_{s_y,\perp} \neq 0$ are tuned up, the behavior near the points $C,D$ versus $F,G$ are qualitatively different. The pairing gap minima near $F,G$ never go to zero as long as $\varphi\neq 0$. In particular, we find:
	\begin{align}
		\Delta_{F,G}=\left\{\begin{array} {cc} \sqrt{(\Delta\cos\frac{\varphi}{2}-b)^2+\Delta^2\cos^2\xi\sin^2\frac{\varphi}{2}}, & b<\Delta\cos\frac{\varphi}{2}\\ \Delta\cos\xi|\sin\frac{\varphi}{2}|,&  b\geq \Delta\cos\frac{\varphi}{2}\end{array}\right.\label{eqn:gap_FG}
	\end{align}
	where
	\begin{align}
		b\equiv
		\begin{cases}
		    |\mu_B H_y\sin\xi -t_{s_y,\perp}| & \mbox{ for point } F,\\
		    |\mu_B H_y\sin\xi + t_{s_y,\perp}| & \mbox{ for point } G.
        \end{cases}
	\end{align}

    On the contrary, near points $C,D$ topological phase transition occurs with four Dirac nodes (two near each point) emerging when $H_y=H^{\text{topo.}}_{c}$, where the critical Zeeman field strength is
	\begin{align}
		\mu_B H^{\mathrm{topo.}}_{c}\equiv\Delta\sqrt{\sec^2\xi\cos^2\frac{\varphi}{2}+\sin^2\frac{\varphi}{2}}.\label{eqn:Hc_topo}
	\end{align}
	When $H_y> H^{\mathrm{topo.}}_{c}$ the four Dirac nodes acquire topological mass gap, transferring a Chern number of 4:
	\begin{align}
		\Delta^{\text{topo.}}_{C,D}=\sin\varphi\frac{\Delta^2\sin^2\xi}{\hbar v_{\text{Ising}}k_F\theta}\sqrt{\left(\frac{H_y}{H^{\text{topo.}}_{c}}\right)^2-1}\label{eqn:topo_gap}
	\end{align}
	The details of the calculations are presented in Appendix \ref{app:analytical}.

	Note that different from $\Delta_{F,G}$, the topological gap $\Delta^{\text{topo.}}_{C,D}$ $\propto\Delta^2$. This is because $\Delta$-linear order gap remains zero in the present linear-order $\bm{k\cdot p}$ effective theory, and the second order perturbation plays the dominant role: the topological phase is always realized when $H_y>H^{\text{topo.}}_{c}$. 

	When a higher order $\bm{k\cdot p}$ expansion is considered, we do find that a non-topological $\Delta$-linear order gap near $C,D$ becomes nonzero (see Appendix \ref{app:analytical}). When this gap is large the topological superconductivity will be destroyed. Based on perturbative calculations, the topological phase requires the following criterion to be satisfied (See Appendix \ref{app:analytical} Eq.(\ref{eqn:v2_condition}) for details):
	\begin{align}
		t_\perp^2 &< \frac{\Delta \hbar k_F v^2_{\text{Ising}}}{v_2}a(\xi,\varphi)\sqrt{\left(\frac{H_y}{H_c^{\text{topo.}}}\right)^2-1},\label{Phase boundary}\\
		a(\xi,\varphi)&\equiv \tan\xi\sin\xi\sqrt{\sec^2\xi\cos^2\frac{\varphi}{2}+\sin^2\frac{\varphi}{2}}.\nonumber
	\end{align}
	Here $v_2$ is a velocity parameter (defined in Eq.(\ref{eqn:v2_definition})) in the quadratic-order $\bm{k\cdot p}$ expansion, $\hbar v^{\text{NbSe}_2}_2=0.94$ and $\hbar v^{\text{TaS}_2}_2=3.1$ eV$\cdot$\AA based on our electronic structure calculations (See Appendix \ref{app:phase_diagram_details}). Namely, $t_\perp$ cannot be too large. This criterion serves as the phase boundary between the trivial and topological phases, and is plotted in FIG.\ref{fig:t-H phase digram} as the dashed black line.

	The previous perturbative regime well captures the Chern-number-12 topological phase. Aiming at understanding the Chern-number-6 topological phase in the numerical phase diagrams, we have performed the analytical calculations in a different perturbative regime: $t_\perp\ll\Delta,\hbar v_{\text{Ising}}k_F\theta\ll\Delta$, and reproduced the Chern-number-6 topological phase (see Appendix \ref{app:analytical}).

\section{Discussion and conclusions}
Before concluding, we would like to remark on a few experiment-related issues. 

\emph{Chiral edge modes:} When topological superconductivity is realized in the proposed heterostructure in Fig.\ref{fig: proposals}(a), the edge of the ferromagnetic buffer layer is the natural boundary between the topological superconductivity and trivial superconductivity. This is because outside the buffer layer region, due to the lack of Zeeman exchange field, gapped trivial superconductivity is realized in the TMD bilayer. Majorana chiral edge modes are then sharply located at this edge, leading to well-known quantized thermal Hall conductance $\frac{\kappa_{xy}}{T}=C\frac{\pi}{12}\frac{k_B^2}{\hbar}$, where $C$ is the Chern number. In addition, the edge modes can detected via scanning tunnelling microscopy as mid-gap states.

Mid-gap states located at the edge of a superconducting material has non-topological explanations, such as the Yu-Shiba-Rusinov bound states. However, in an appropriate regime, as shown in Fig.\ref{fig:theta-phi phase digram}, we note that a single device may realize trivial to topological phase transition while the Josephson phase $\varphi$ is tuned up. Such a $\varphi$-driven topological phase transition has unique experimental signatures since the majorana edge states are expected to exist only in the topological phase, which may be used to sharply identify the nature of the mid-gap states. 

	\vspace{1em}

\emph{The effect of in-plane external magnetic field:} When an in-plane magnetic field is applied to the proposed heterostructures, the Josephson phase $\varphi(x)=2\pi x/L$ becomes spatial dependent along the in-plane transverse direction, where $L=\frac{\Phi_0}{H_{\text{ext.}}d}$ and $d$ is the effective thickness of the junction. For instance, an external magnetic field $H_{ext.}\sim 0.5$T is needed to reorient the magnetic moment of CrBr$_3$ to an in-plane direction, corresponding to $L\sim 2\mu$m if $d\sim 20\text{\AA}$ is used. Due to the fact that the Chern number flips sign when $\varphi\rightarrow -\varphi$, the topological superconductivity is expect to form spatial stripes, or domains with width $L/2$, with alternating Chern numbers, e.g., $C=\pm12$ in which case 24 chiral majorana states are expected to form at each domain-wall. In Appendix \ref{app:domain_wall_modes} we estimate the spatial spread $l_\perp$  of the domain-wall majorana states along the transverse direction of the domain-wall. Only when $l_\perp$ is much smaller than the stripe width $L/2$ are the domain-wall states well-defined. We find that for a generic magnetic field direction $l_\perp$ is comparable with $L/2$ for $H_{ext.}\sim 0.5$T. However, when $H_{ext.}$ is parallel to one $\Gamma-M$ direction, 8 among the 24 domain-wall majorana states have an $l_\perp$ that can be $5\sim 10$ times smaller than $L/2$. These domain wall chiral states may be observable in probes such as STM or thermal transport.

	\vspace{1em}
	\emph{Charge-density-wave Order:} It is known that Ising superconductivity in 2H-$\mathrm{NbSe_2}$ or 2H-$\mathrm{TaS_2}$ coexists with the charge-density-wave (CDW) order \cite{ugeda2016characterization,xi2015strongly,nakata2018anisotropic}. For the Fermi pockets around $\Gamma$, the Fermi surface folding due to CDW occurs near the $\Gamma$-$K$ direction, which would not qualitatively affect the low energy physics near the $\Gamma$-$M$ direction, which we have been focusing on in this paper. 

	\vspace{1em}
	\emph{Rashba Spin-orbit Coupling:} As emphasized before, the $z\rightarrow-z$ mirror symmetry forbids the Rashba spin-orbit coupling in the monolayer TMD. A Rashba spin-orbit coupling breaks an important invariance of Eq.(\ref{eqn:bilayer_bcs}) related to the combination of particle-hole and time-reversal transformation $\mathsf{PH}\circ\mathsf{TR}=\sigma_y\tau_x$. $\mathsf{PH}\circ\mathsf{TR}$ always sends $h_{\text{BCS}}^{\text{bilayer}}(\bm{k},H_y,\varphi)\mapsto -h_{\text{BCS}}^{\text{bilayer}}(\bm{k},-H_y,-\varphi)$. In the absence of the Rashba spin-orbit coupling one may flip the sign of $H_y$ and $\varphi$ by a complex conjugation since all other terms in the Hamiltonian are real: $-h_{\text{BCS}}^{\text{bilayer}}(\bm{k},-H_y,-\varphi)=-(h_{\text{BCS}}^{\text{bilayer}})^*(\bm{k},H_y,\varphi)$. Namely there is an invariance protecting the Bogoliubov quasiparticle spectrum being $\pm E$ symmetric at any $\bm k$. Such an invariance is lost in the presence of a Rashba spin-orbit coupling, leading to Bogoliubov Fermi pockets \cite{shaffer2020crystalline}. 
	
	In the proposed heterostructures, despite the lack of the $z\rightarrow-z$ mirror symmetry, the intrinsic Rashba coupling induced by the vdW interlayer interactions can be safely neglected since it is extremely weak (see in Appendix \ref{app:phase_diagram_details}). Although an extrinsic, substrate-induced Rashba coupling is possible, in this paper we do not consider this effect, since nevertheless this coupling could be tuned to zero via gating.
	
    \vspace{1em}
	\emph{Interlayer Tunneling Strength:} It is clear from the phase diagrams in FIG. \ref{fig:t-H phase digram} and criterion Eq.(\ref{Phase boundary}) that the interlayer tunneling $t_\perp$ cannot be too large in order to realize the topological superconductivity ($<10$meV). On the other hand, based on our DFT calculation (see Appendix \ref{app:phase_diagram_details}), due to the $d_{z^2}$ nature of Fermi pockets around $\Gamma$-point, a fairly large interlayer tunneling of the bilayer systems of $\mathrm{NbSe_2}$ or $\mathrm{TaS_2}$ \emph{without} a buffer layer is found near the point-$P$ ($>50$meV) \footnote{In literature \protect\url{https://www.nature.com/articles/s41467-018-03888-4}, a smaller value of tunneling about 15meV for bilayer $\mathrm{NbSe_2}$ is reported. We ascribe such discrepancy to the choice of van der Waals pseudopotentials.}. This actually motivates us to consider the insulating buffer layer in the proposed heterostructures. In Appendix \ref{app:phase_diagram_details}, as an estimate, we performed DFT calculations on a $\mathrm{TaS_2}/\mathrm{WS_2}/\mathrm{TaS_2}$ heterostructure, and $t_\perp\sim 5$meV is found -- well within the regime where topological superconductivity is realized.
	
	\vspace{1em}
	In summary, we theoretically propose twisted bilayer vdW Ising superconductors as a new flexible and tunable platform to realize chiral topological superconductivity with Chern numbers. In the simplest setup, an insulating buffer layer with an in-plane ferromagnetic moment is introduced between the TMD Ising superconductors such as $\mathrm{NbSe_2}$ or $\mathrm{TaS_2}$, providing Zeeman exchange field in the TMD layers via the magnetic proximity effect. We show that the out-of-plane supercurrent induces topological superconductivity over a large parameter regime, and the characteristic majorana chiral edge modes are localized on the edge of the ferromagnetic buffer layer. We hope that the present study may motivate further experimental and theoretical investigations on such vdW heterostructures.
	
	\vspace{1em}
	\emph{Acknowledgement} Y.R. and X.H. acknowledge the support from National Science Foundation under Grant No. DMR-1712128. We thank Di Xiao for helpful discussions. X.H. thanks the open discussion of distributive computation on \url{Julia Discourse}, and the HPC resources of \emph{Andromeda Cluster} at Boston College.

\appendix
\section{Domain-wall chiral modes}\label{app:domain_wall_modes}
	Assuming inter-node scattering being weak, here we may use the Dirac equation for a single node to estimate the spatial size of the chiral domain-wall modes. Denoting the magnetic field direction as $\hat B$, we have the spatial-dependent Josephson phase $\varphi(\bm r)=\frac{2\pi}{L}\hat B\times\hat z  \cdot \bm r\equiv \frac{2\pi}{L}\hat n\cdot \bm r$, where we defined unit vector $\hat n\equiv \hat B\times \hat z$. Based on Eq.(\ref{eqn:topo_gap}), the spatial-dependent Dirac equation becomes:
	\begin{align}
		H=v_x p_x\Sigma_x+v_y p_y\Sigma_z+m_0\sin\varphi(\bm r)\Sigma_y,
	\end{align}
	where
	\begin{align}
		m_0\equiv\frac{\Delta^2\sin^2\xi}{\hbar v_{\text{Ising}}k_F\theta}\sqrt{\Big(\frac{H_y}{H^{\text{topo.}}_{c}}\Big)^2-1}.
	\end{align}
	$v_x=v_F\cos\theta^+$ and $v_y=v_{\text{Ising}}\sin\theta^+$ according to Eq.(\ref{heff for P+ and P-}), and $v_x\gg v_y$ since $\theta^+$ is an angle parameter of order unity. 
	The mass-sign-changing domain walls are located at $\hat n\cdot \bm r=\frac{k L}{2}$, and $k\in\mathbb{Z}$. Near each domain wall, we may linearize the mass as $m_0\sin\varphi(\bm r)\sim (-1)^k m_0\tilde \varphi(\delta \bm r)$, where $\tilde\varphi(\delta \bm r)=\varphi(\bm r)-k\pi=\frac{2\pi\hat n\cdot \delta\bm r}{L}$, and $\delta \bm r$ is the position measured from the domain wall.

	It is now convenient to rotate into the coordinate system with axes $(x_\parallel,x_\perp)$ along and perpendicular to the domain-wall. The corresponding momentum are denoted as $(p_\parallel,p_\perp)$. Writing the cosine $c$ and sine $s$ of the rotation angle between the two coordinate systems, then $p_x\equiv c p_\parallel- s p_\perp$ and $p_y\equiv s p_\parallel+c p_\perp$, and we have
	\begin{align}
		H&=v_x (c p_\parallel- s p_\perp)\Sigma_x+v_y (s p_\parallel+c p_\perp)\Sigma_z\nonumber\\
		&\qquad+(-1)^k m_0\frac{2\pi x_\perp}{L}\Sigma_y,
	\end{align}
	$H^2$ has a simple form:
	\begin{align}
		H^2&=v_x^2(c p_\parallel- s p_\perp)^2+v_y^2(s p_\parallel+c p_\perp)^2+m_0^2\left(\frac{2\pi x_\perp}{L}\right)^2\notag\\
		&-(-1)^k \hbar m_0\frac{2\pi}{L}(sv_x\Sigma_z+c v_y\Sigma_x)
	\end{align}
	Since $p_\parallel$ is a good quantum number we are left with a one-dimensional harmonic oscillator involving $p_\perp,x_\perp$:
	\begin{align}
		H^2&=\big((cv_x)^2+(sv_y)^2\big)p_\parallel^2+\big((sv_x)^2+(cv_y)^2\big)(p_\perp - p_0(p_\parallel) )^2\notag\\
		+&m_0^2\left(\frac{2\pi x_\perp}{L}\right)^2-(-1)^k \hbar m_0\frac{2\pi}{L}(sv_x\Sigma_z+c v_y\Sigma_x),
	\end{align}
	where $p_0$ is a linear function of $p_\parallel$. The corresponding mass and frequency for this harmonic oscillator are:
	\begin{align}
		M&=\frac{1}{2}[(sv_x)^2+(cv_y)^2]^{-1}\notag\\
		\omega&=\sqrt{2M^{-1}}\frac{m_02\pi}{L}.
	\end{align}
	The energy levels are:
	\begin{align}
		E_n=2\hbar\sqrt{(sv_x)^2+(cv_y)^2}\frac{m_02\pi}{L}(n+\frac{1}{2}).
	\end{align}
	The zero energy state of $H$ and $H^2$ for $p_\parallel=0$, corresponding to the chiral majorana modes, is the $n=0$ ground state of the harmonic ocsillator. The last term in $H^2$ chooses a specific eigenstate of $(sv_x\Sigma_z+c v_y\Sigma_x)$, so that the zero-point energy in the harmonic oscillator is exactly canceled. Since we are working with complex fermion, this zero energy state corresponds to two chiral majorana modes.

	The spatial spread $l$ of the chiral model along the $x_\perp$ direction can be readily read out from the harmonic oscillator ground state $\psi(x_\perp)\propto e^{-x_\perp^2/(2l_\perp^2)}$:
	\begin{align}
		l_\perp&=\sqrt{\frac{\hbar}{M\omega}}=\sqrt{\frac{\hbar [(sv_x)^2+(cv_y)^2]^{1/2} L}{2\pi m_0}}
	\end{align}
	
	$l_\perp$ is proportional to the square root of the velocity $[(sv_x)^2+(cv_y)^2]^{1/2}$. Note that $v_x\gg v_y$, for a generic magnetic field direction, $l_\perp\sim \sqrt{\frac{\hbar v_x L}{2\pi m_0}}$. Assuming $\hbar v_x\sim 1\mathrm{eV}\cdot\text{\AA}$, $m_0\sim 0.1\mathrm{meV}$, and $L\sim 2\mathrm{\mu m}$, one finds $l_\perp\sim 0.4\cdot L/2$.
	
    To minimize $l_\perp$, one may choose the magnetic field direction (i.e., the domain-wall direction) being along $x$ (i.e., one $\Gamma$-$M$ direction), leading to $l_\perp=\sqrt{\frac{\hbar v_y L}{2\pi m_0}}$ for 4 nodes near this $\Gamma$-$M$ direction among the total of 12 nodes. In this case, among the 24 majorana domain wall states, 8 of them have $l_\perp\sim 0.2\cdot L/2$ (for TaS$_2$) and $l_\perp\sim 0.1\cdot L/2$ (for NbSe$_2$) using the parameters above. 

\section{A mean-field treatment for the ferromagnetic buffer layer}\label{app:magnetic_buffer_layer_perturbation}
	We will show in this section that, for the proposed setup in Fig.\ref{fig: proposals}(a) with a ferromagnetic buffer layer, the direct couplings between the top/bottom layer and the middle layer will induce an in-plane exchange Zeeman-field $H_{\text{exch.}}$, and a spin-dependent tunneling $t_{s,\perp}$. In a simple perturbative mean-field treatment, $\mu_B H_{\text{exch.}}=t_{s,\perp}$.

	Without loss of generality, we can work in the basis where the buffer layer Hamiltonian is diagonalized $\bm{h}^\text{m}=\mathop{\mathrm{diag}}\{h^{\text{m}}_\uparrow, h^\text{m}_\downarrow\}$, where $\uparrow$ and $\downarrow$ indicate the spin components parallel and antiparallel to \emph{the in-plane ferromagnetic moment direction}. For simplicity we assume $h^{\text{m}}_\uparrow$, $h^{\text{m}}_\downarrow$ each contains a single band, which can be easily generalized to multi-band cases.

	In the absence of the middle layer, we consider the spinful TMD bilayer Hamiltonian $\bm{h}^\text{bilayer}=\mathop{\mathrm{diag}}\{\bm{h}^\text{t}(\bm{k}), \bm{h}^\text{b}(\bm{k})\}$. The full Hamiltonian for the trilayer system is then:
	\begin{equation}\label{trilayer Hamiltonian}
		H^{\text{trilayer}}(\bm{k})=\left(\begin{array}{c|c}
			\begin{matrix}
				\bm{h}^\text{t}(\bm{k}) & \\
				 & \bm{h}^\text{b}(\bm{k})
			\end{matrix} & \text{\Large $\bm{V}$} \\
			\hline
			\text{\Large $\bm{V}^T$} & \begin{matrix}
				h^\text{m}_\uparrow & \\
				 & h^\text{m}_\downarrow
			\end{matrix}
		\end{array}\right),
	\end{equation}
	where $\bm{V}=(\bm{V}_1,\bm{V}_2)$, $\bm{V}_1=(t^\text{t}_\uparrow,0,t^\text{b}_\uparrow,0)^T$, and $\bm{V}_2=(0,t^\text{t}_\downarrow,0,t^\text{b}_\downarrow)^T$.\par
	Due to the $z\rightarrow -z$ mirror symmetry for the TMD bilayer at the zero twisting angle, one may assume these direct hoppings to be layer-independent $t^{\text{t,b}}_{\uparrow,\downarrow}=t_{\uparrow,\downarrow}$ at small twisting angles. Since the middle buffer layer is an insulator, 	$h^\text{m}_\uparrow,	h^\text{m}_\uparrow$ are high energy states. Standard perturbation theory gives
	\begin{align}
		&H^{\text{bilayer}}_\text{induced}(\bm{k})\simeq \bm{h}^{\text{bilayer}}-\sum_{i=1,2} \bm{V}_i(\bm{h}^\text{m})^{-1}\bm{V}_i^T\nonumber\\
		&=\bm{h}^{\text{bilayer}}-\dfrac{t_\uparrow^2}{h^\text{m}_\uparrow}\left(\begin{array}{cccc}
			1 & 0 & 1 & 0\\
			0 & 0 & 0 & 0\\
			1 & 0 & 1 & 0\\
			0 & 0 & 0 & 0
		\end{array}\right)-\dfrac{t_\downarrow^2}{h^\text{m}_\downarrow}\left(\begin{array}{cccc}
			0 & 0 & 0 & 0\\
			0 & 1 & 0 & 1\\
			0 & 0 & 0 & 0\\
			0 & 1 & 0 & 1
		\end{array}\right)\nonumber\\
		&=\bm{h}^{\text{bilayer}}-\delta\mu\sigma_0\nu_0+t_\perp\sigma_0\nu_x+\mu_B H_{\text{exch.},z}\sigma_z\nu_0+t_{s,\perp}\sigma_z\nu_x,\label{proximity-induced Heff}
	\end{align}
	where we separate out the second-order perturbation of bilayer Hamiltonian into four parts: the chemical potential shift $\delta\mu$, the spin-independent tunneling $t_\perp$, the (proximity-induced) Zeeman field $H_{\text{exch.}}$, and the spin-dependent tunneling $t_{s,\perp}$, where. 
	\begin{align}
		-\delta\mu& = t_\perp=-\dfrac{1}{2}\left(\dfrac{t^2_\uparrow}{h^\text{m}_\uparrow}+\dfrac{t^2_\downarrow}{h^\text{m}_\downarrow}\right),\label{proximity-induced constraint: mu and t_perp}\\
		\mu_BH_{\text{exch.}}&=t_{s,\perp}=-\dfrac{1}{2}\left(\dfrac{t^2_\uparrow}{h^\text{m}_\uparrow}-\dfrac{t^2_\downarrow}{h^\text{m}_\downarrow}\right).\label{proximity-induced constraint: h_exch and t_s,perp}
	\end{align}
	
\section{Details of the phase diagram calculations and discussions on the  stacking-dependence}\label{app:phase_diagram_details}
	All phase diagrams of the proposed heterostructures involving either 2H-NbSe$_2$ or 2H-TaS$_2$ are obtained by listing the gap size for all eight nodes (topological and trivial) and picking out the \emph{minima} of them. To capture the physics among entire phase-diagrams (particularly the large-$H_\text{exch.}$ regions), we follow Ref.\cite{xi2016ising} to introduce the pair breaking equation throughout our numerical calculation
	\begin{equation*}
		\ln(T_c/T_{c0})+\psi\left(\dfrac{1}{2}+\dfrac{\mu_B H^2/H_{\text{so}}}{2\pi k_B T_c}\right)-\psi(1/2)\equiv0,
	\end{equation*}
	where $\psi$ is the digamma function, $T_{c0}$ is the zero-field critical temperature ($3.0$K for $\mathrm{NbSe_2}$ \cite{xi2016ising} and $3.4$K for $\mathrm{TaS_2}$ \cite{yang2018enhanced}), and $H_{\text{so}}$ can be taken as a fitting parameter to match the asymptotic Ginzberg-Landau behavior $H\sim H_0\sqrt{1-T/T_{c0}}$ for $T_c\sim T_{c0}$, where $H_0\sim\sqrt{H_{\text{so}}H_p}$ and the Pauli limit  $H_p=\frac{\Delta_0}{\sqrt{2}\mu_B}=(1.86~\mathrm{T/K})\cdot T_{c0}$ assuming the $g$-factor equals to 2 and the BCS formula $\Delta_0=1.76k_B T_{c0}$ still to hold \footnote{For few-layer $\mathrm{TaS_2}$, the relation is also reported to be $\Delta_0=2.2k_B T_{c0}$ in \protect\url{https://www.nature.com/articles/s41563-018-0061-1}. Since such coefficient is material- and stacking-dependent, we will still take the value of BCS approximation $\Delta_0=1.76k_B T_{c0}$ as a demonstration.}. The best fitting of experimental data gives $H_0=43.6\mathrm{T}$ for $\mathrm{NbSe_2}$ and $H_0=65.6\mathrm{T}$ for $\mathrm{TaS_2}$ \cite{de2018tuning}, so we take $H_{\text{so}}^{\mathrm{NbSe_2}}=340\mathrm{T}$ and $H_{\text{so}}^{\mathrm{TaS_2}}=680\mathrm{T}$ throughout our numerical calculations.\par
	All parameters we input in our numerical calculations of phase diagrams (up to fourth-order of $\bm{k\cdot p}$ expansion) are extracted from the monolayer tight-binding models obtained from LDA calculation by QUANTUM ESPRESSO \cite{giannozzi2009quantum,giannozzi2017advanced} and Wannier90 routine \cite{pizzi2020wannier90}. Our strategies is to first build a AA-stacking $\beta$-bilayer slab system for both materials, then relax the structure with non-local van de Waals correlation functionals to get the proper layer distance and (spin-independent) tunneling strengths. For convergence reasons, we choose to use the functional vdW-DF-C6 \cite{tkatchenko2009accurate,berland2019van} for $\mathrm{TaS_2}$ while vdW-DF2-C09 \cite{lee2010higher,cooper2010van} for $\mathrm{NbSe_2}$. The resulting relaxed layer separation is $d_\perp^{\mathrm{TaS_2}}=6.57\text{\AA}$ and $d_\perp^{\mathrm{NbSe_2}}=6.80\text{\AA}$. Since tunneling processes only occur around the interesction points $P$, with their strengths to be half of the energy splitting around the Fermi energy $\varepsilon_F$, we can directly read them out from the the non self-consistent output at $\bm{k}=(k_F,0)$, reading $t_\perp^{\mathrm{TaS_2}}=56.9\mathrm{meV}$ and $t_\perp^{\mathrm{NbSe_2}}=116.2\mathrm{meV}$.\par
	Next, we take the relaxed structure as an input to build a monolayer slab system and perform a LDA calculation to obtain a monolayer electronic structure. Note that this monolayer relaxed structure weakly breaks the $z\rightarrow -z$ symmetry, but preserves the mirror planes parallel with the $z$-axis. The LDA result is used to fit the $\bm{k\cdot p}$ model up to quartic order. More precisely, we fit the model consistent with the $y\rightarrow -y$ mirror symmetry:
	\begin{align}\label{monolayer quartic kdotp expansion}
		H^{\text{mono}}_{\text{quartic}}(\bm{k})&=\bigg[\varepsilon_F +v_F k_x + a k_y^2 + p_5 k_x^2 + p_6 k_x k_y^2\nonumber\\
		&\quad + p_7 k_y^4 + p_8 k_x^3 + p_9 k_x^4 + p_{10} k_x^2 k_y^2\bigg]\sigma_0\nonumber\\
		&\quad + p_{11}\sigma_y +\bigg[ v_{\text{Ising}} k_y + p_{13} k_x k_y + p_{14} k_y^3\nonumber\\
		&\quad + p_{15} k_x^2 k_y + p_{16} k_x^3 k_y + p_{17} k_x k_y^3\bigg]\sigma_z.
	\end{align}
	The results for both $\mathrm{TaS_2}$ and $\mathrm{NbSe2}$ are listed as following: (in the eV-\AA~unit system and we set $\hbar=1$, as a complement of the TABLE.\ref{table:parameters} given in the main text)
	\begin{widetext}
		\begin{center}
			\begin{tabular}{c|ccccccccccccccccc}
				\hline\hline
				\diagbox{Materials}{Coef. $p_i$}  & $k_F$ & $v_F$ & $a$ & $p_5$ & $p_6$ & $p_7$ & $p_8$ & $p_9$ & $p_{10}$ & $p_{11}$ & $v_{\text{Ising}}$ & $p_{13}$ & $p_{14}$ & $p_{15}$ & $p_{16}$ & $p_{17}$\\
				\hline
				$\mathrm{TaS_2}$  & 0.54 & -2.89 & 0.25 & 2.34 & 17.82 & -8.72 & 11.53 & -8.67 & 17.20 & 0.00 & 0.36 & 1.41 & -1.06 & -0.72 & -2.11 & -8.45\\
				\hline
				$\mathrm{NbSe_2}$ & 0.48 & -2.22 & -0.76 & 1.36 & 15.10 & -2.98 & 9.85 & 0.76 & 26.15 & 0.00 & 0.12 & 0.86 & -0.20 & 0.25 & -3.56 & -4.14\\
				\hline\hline
			\end{tabular}	
		\end{center}
	\end{widetext}
    The intrinsic Rashba SOC, ignored in Eq.\eqref{eqn:1layer}, appears as the vanishingly small $p_{11}$ in the above fitting. Due to the loss of the $z\rightarrow -z$ mirror symmetry in the vdW relaxed structure, in principle this coupling may be nonzero. However, we find that the raw energy splittings (corresponding to $2p_{11}$) at the point-$P$ is less than $10^{-10}\text{meV}$ for both $\mathrm{NbSe_2}$ and $\mathrm{TaS_2}$, and conclude that weak vdW interactions cannot lead to any sizable intrinsic Rashba SOC. 
    
    The values of aforementioned spin-independent tunnelings $t_\perp$ for bilayer $\mathrm{NbSe_2}$ and $\mathrm{TaS_2}$ are far beyond the topological superconducting regime that we find in the numerical phase diagrams. That actually motivates us to propose the buffer layer heterostructures in FIG.~\ref{fig: proposals}. To have an estimation on the magnitude of the spin-independent tunneling $t_\perp$ \emph{after} the insertion of an insulating buffer layer, we build a trilayer slab system $\mathrm{TaS_2}/\mathrm{WS_2}/\mathrm{TaS_2}$ of ABA and AAA stackings. Since $\mathrm{2H}$-$\mathrm{TaS_2}$ and $\mathrm{2H}$-$\mathrm{WS_2}$ have similar lattice constants, we use the lattice constants of $\mathrm{2H}$-$\mathrm{TaS_2}$ for the trilayer system without enlarging the supercell \footnote{We use the crystalline data from Spinger Materials, see \protect\url{https://materials.springer.com/isp/crystallographic/docs/sd_0457067} and \protect\url{https://materials.springer.com/isp/crystallographic/docs/sd_0551013}.} ($a=0.331$nm and $c=0.121$nm for $\mathrm{2H}$-$\mathrm{TaS_2}$, and $a=0.315$nm and $c=0.121$nm for $\mathrm{2H}$-$\mathrm{WS_2}$). We then again relax the positions for all atoms of the trilayer system with the non-local van der Waals functional vdW-DF-C6 \cite{tkatchenko2009accurate,berland2019van} along the out-of-plane direction. We find, after an inserting of a non-magnetic buffer layer $\mathrm{WS_2}$, the separation between top and bottom $\mathrm{TaS_2}$ layers for both stackings almost doubles to $d_\perp^{\mathrm{TaS_2},\text{ABA}}=12.00\text{\AA}$ and  $d_\perp^{\mathrm{TaS_2},\text{AAA}}=13.18\text{\AA}$, and the corresponding strengths of the spin-independent tunneling reduce to $t_\perp^{\mathrm{TaS_2},\text{ABA}}=5.0\mathrm{meV}$ and $t_\perp^{\mathrm{TaS_2},\text{AAA}}=4.5\mathrm{meV}$.

\section{Details of analytical perturbative calculations}\label{app:analytical}
	We will work with the atomic units by default throughout the derivation here. For example, $\hbar=\mu_B=1$.
	\hfill\\
	\subsection{Perturbative theory for $\sqrt{(v_{\text{Ising}}k_F\theta)^2+4t_\perp^2}\gg \Delta $: $C=12$ Topological Phases}
		In this section, we will give a perturbative analysis on the twisted bilayer Hamiltonian given in the main text. We will focus on the regime when $\sqrt{(v_{\text{Ising}}k_F\theta)^2+4t_\perp^2}\gg \Delta $ and give an explanation on the origin of $C=\pm12$ topological phases. Mass gaps for all nodes and the phase boundary will also be also derived.
		\subsubsection{Construction of Hamiltonian}
			Let us start with constructing the effective Hamiltonian. We will consider a small twisting angle $\theta$, and keep the $\bm{k\cdot p}$ expansion up to $\mathcal{O}(\theta^2)$ (the reason will be given in the next subsection). As is discussed in the main text, in-plane mirror symmetry prohibit the existence of Rashba SOC, so the monolayer effective Hamiltonian around the intersection point $P$ of the Fermi surface and the $\Gamma$-$M$ line simply reads
			\begin{equation}\label{monolayer Hamiltonian}
				h^{\text{mono}}(\bm{k})=(v_F k_x + a k_y^2)+v_{\text{Ising}} k_y \sigma_z,
			\end{equation}
			where $a$ is the coefficient of $\bm{k\cdot p}$ expansion. We keep a quadratic term $ak_y^2$ here because it turns out that this term contributes to the leading order deviation from the linear $\bm{k\cdot p}$ expansion. Equation \eqref{monolayer Hamiltonian} can be easily extend to higher orders and we did use the fitting fourth-order $\bm{k\cdot p}$ in our numerical calculations, see section V.\par
			Without loss of generality, we can consider a bilayer system with the top layer rotated by $\theta/2$, and the bottom layer rotated by $-\theta/2$. This is achieved by simple replacement $k_x\mapsto(k_x+k_F)\cos\frac{\theta}{2}\pm k_y\sin\frac{\theta}{2}-k_F$ and $k_y\mapsto\mp(k_x+k_F)\sin\frac{\theta}{2}+k_y\cos\frac{\theta}{2}$ and expansion still up to $\theta^2$. The resulting bilayer Hamiltonian \emph{without} tunneling is then
			\begin{widetext}
				\begin{align}\label{bilayer Hamiltonian without Tunneling}
					h^{\text{bilayer without tunneling}}(\bm{k})&=\left(v_F k_x+ak_y^2+\dfrac{ak_F^2}{4}\theta^2-\dfrac{v_F k_F}{8}\theta^2\right)+\left(-ak_F k_y \theta+\dfrac{\theta}{2}v_F k_y\right)\nu_z+v_{\text{Ising}} k_y\sigma_z-v_{\text{Ising}} k_F\dfrac{\theta}{2}\sigma_z\nu_z \nonumber\\
					&=\left(v_F k_x+ak_y^2-\dfrac{v_2 k_F}{8}\theta^2\right)+\dfrac{\theta}{2}v_2k_y\nu_z+v_{\text{Ising}} k_y\sigma_z-\dfrac{\theta}{2}v_{\text{Ising}} k_F\sigma_z\nu_z,
				\end{align}
			\end{widetext}
			where $\nu_i$ are Pauli matrices within the layer space and we defined
			\begin{equation}
				v_2\equiv v_F-2ak_F.\label{eqn:v2_definition}
			\end{equation}
			In Eq.(\ref{bilayer Hamiltonian without Tunneling}), the crucial role played by $v_2$ is due to the second term, which gives the leading order correction to the wavefunctions comparing with the linear $\bm{k\cdot p}$ expansion.
			\indent Next, let us try to add interlayer tunnelings to \eqref{bilayer Hamiltonian without Tunneling}. Based on \emph{two-center approximation} \cite{bistritzer2011moire}, it can be shown that (see in Appendix \ref{app:magnetic_buffer_layer_perturbation}) the interlayer tunneling have a weak spatial-dependence and will be considered as a constant here $t=t_\perp$ (we will consider the spin-dependent tunnelings $t_{s,\perp}$ later). We thus have
			\begin{align}\label{bilayer Hamiltonian}
				h^{\text{bilayer}}&=\left(v_F k_x+ak_y^2-\dfrac{v_2k_F^2}{8}\theta^2\right)\nonumber\\
				&\quad+\dfrac{\theta}{2}v_2 k_y\nu_z+v_{\text{Ising}} k_y\sigma_z-\dfrac{\theta}{2}v_{\text{Ising}} k_F\sigma_z\nu_z+t\nu_x.
			\end{align}
			Hamiltonian \eqref{bilayer Hamiltonian} reduce to simple two-band model \cite{bernevig2013topological} (so can be readily diagonalized) for $\sigma_z=\pm1$. Keeping up to $\mathcal{O}(\theta^2)$, we have, for $\sigma_z=+1$,
			\begin{align*}
				\varepsilon_\pm^\uparrow &=\left(v_F k_x +a k_y^2-\dfrac{\theta^2}{8}v_2 k_F\right)+v_{\text{Ising}}k_y\pm\dfrac{\Gamma}{2}\left[1-\dfrac{v_2\cdot\cos^2\xi}{v_{\text{Ising}} k_F}k_y\right]
			\end{align*}
			and for $\sigma_z=-1$,
			\begin{equation*}
				\varepsilon^\downarrow_\pm=\left(v_F k_x +a k_y^2-\dfrac{\theta^2}{8}v_2 k_F\right)-v_{\text{Ising}}k_y\pm\dfrac{\Gamma}{2}\left[1+\dfrac{v_2\cdot\cos^2\xi}{v_{\text{Ising}} k_F}k_y\right],
			\end{equation*}
			where
			\begin{equation*}
				\xi\equiv \arctan\dfrac{2t}{v_{\text{Ising}} k_F \theta},\quad\text{and}\quad\Gamma\equiv\sqrt{(v_{\text{Ising}}k_F\theta)^2+4t^2}.
			\end{equation*}

			There are some crossing points between the spin-up and spin-down Fermi surfaces. Solving $\varepsilon_\pm^\uparrow=\varepsilon_\pm^\downarrow=0$, we find two such points $F$ and $G$ along the $k_x$ direction ($\Gamma$-$M$ line)
			\begin{equation}\label{F,G coordinates withouth Delta and phi}
				F=\left(-\dfrac{\Gamma}{2v_F}+\dfrac{v_2k_F\theta^2}{8v_F}, 0\right),\quad G=\left(+\dfrac{\Gamma}{2v_F}+\dfrac{v_2k_F\theta^2}{8v_F}, 0\right).
			\end{equation}
			At $F$, the two low energy bands come from $\varepsilon_-^{\uparrow,\downarrow}$, while at $G$, the two low energy bands are formed by $\varepsilon_+^{\uparrow,\downarrow}$. Interestingly, there are another two such points $C$ and $D$ emergent along the $k_y$-direction that have different low-energy bands information. We also have
			\begin{align}
				C&=\left(-\dfrac{k_F\theta^2}{8}-\dfrac{a t^2}{v_Fv_{\text{Ising}}^2}, -\dfrac{\Gamma}{2v_{\text{Ising}}}\right),\label{C coordinates withouth Delta and phi}\\
				D&=\left(-\dfrac{k_F\theta^2}{8}-\dfrac{a t^2}{v_Fv_{\text{Ising}}^2}, +\dfrac{\Gamma}{2v_{\text{Ising}}}\right).\label{D coordinates withouth Delta and phi}
			\end{align}
			At $C$, the two low-energy bands are formed by $\varepsilon_-^\uparrow,\varepsilon_+^\downarrow$, while at $D$, the two low-energy bands are formed by $\varepsilon_+^\uparrow,\varepsilon_-^\downarrow$. They are all illustrated in the main text. In the following derivation, we will denote the $\bm{k}$-space coordinates of the intersection points $F$ as $\bm{F}_+$, $G$ as $\bm{F}_-$, $C$ as $\bm{C}_-$, and $D$ as $\bm{C}_+$.\par
			The low-energy effective Hamiltonian of each node is then the projection of \eqref{bilayer Hamiltonian} onto these bands. First of all, we need to read out the eigenstates in $\nu$-space. Introducing the elevation angle
			\begin{align*}
				\xi_\pm&=\arctan\dfrac{t}{\frac\theta 2(v_2k_y-v_{\text{Ising}} k_F)}\\
				&\simeq\arctan\left[\left(1\pm\dfrac{v_2k_y}{v_{\text{Ising}} k_F}\right)\dfrac{2t}{v_{\text{Ising}} k_F \theta}\right],
			\end{align*}
		    we obtain:
			\begin{align*}
				\sigma_z=+1:&\quad|\nu_1\rangle=\sin\frac{\xi_+}{2}|\nu_\uparrow\rangle+\cos\frac{\xi_+}{2}|\nu_\downarrow\rangle,\\
				&\quad|\nu_2\rangle=\cos\frac{\xi_+}{2}|\nu_\uparrow\rangle-\sin\frac{\xi_+}{2}|\nu_\downarrow\rangle,\\
				\sigma_z=-1:&\quad|\nu_3\rangle=\cos\frac{\xi_-}{2}|\nu_\uparrow\rangle+\sin\frac{\xi_-}{2}|\nu_\downarrow\rangle,\\
				&\quad|\nu_4\rangle=-\sin\frac{\xi_-}{2}|\nu_\uparrow\rangle+\cos\frac{\xi_-}{2}|\nu_\downarrow\rangle.
			\end{align*}
			We can then lift \eqref{bilayer Hamiltonian} into Nambu representation
			\begin{widetext}
				\begin{equation}\label{bilayer Hamiltonian in Nambu}
					H^{\text{bilayer}}=\left(v_F k_x+ak_y^2-\dfrac{v_2 k_F}{8}\theta^2\right)\tau_z+\dfrac{\theta}{2}v_2k_y\nu_z\tau_z+v_{\text{Ising}} k_y\sigma_z-\dfrac{\theta}{2}v_{\text{Ising}} k_F\sigma_z\nu_z+t\nu_x\tau_z,
				\end{equation}
				and perform the projection. Here $\nu_i$ are Pauli matrices for the Nambu space.\par
				Choosing the basis as following
				\begin{align*}
					\text{At }G:&\quad \{|\tau_\uparrow,\sigma_\uparrow,\nu_2\rangle,|\tau_\uparrow,\sigma_\downarrow,\nu_4\rangle,|\tau_\downarrow,\sigma_\uparrow,\nu_4\rangle,|\tau_\downarrow,\sigma_\downarrow,\nu_2\rangle, |\tau_\uparrow,\sigma_\uparrow,\nu_1\rangle,|\tau_\uparrow,\sigma_\downarrow,\nu_3\rangle,|\tau_\downarrow,\sigma_\uparrow,\nu_3\rangle,|\tau_\downarrow,\sigma_\downarrow,\nu_1\rangle\},\\
					\text{At }F:&\quad \{|\tau_\uparrow,\sigma_\uparrow,\nu_1\rangle,|\tau_\uparrow,\sigma_\downarrow,\nu_3\rangle,|\tau_\downarrow,\sigma_\uparrow,\nu_3\rangle,|\tau_\downarrow,\sigma_\downarrow,\nu_1\rangle,|\tau_\uparrow,\sigma_\uparrow,\nu_2\rangle,|\tau_\uparrow,\sigma_\downarrow,\nu_4\rangle,|\tau_\downarrow,\sigma_\uparrow,\nu_4\rangle,|\tau_\downarrow,\sigma_\downarrow,\nu_2\rangle\},\\
					\text{At }D:&\quad \{|\tau_\uparrow,\sigma_\uparrow,\nu_2\rangle,|\tau_\uparrow,\sigma_\downarrow,\nu_3\rangle,|\tau_\downarrow,\sigma_\uparrow,\nu_3\rangle,|\tau_\downarrow,\sigma_\downarrow,\nu_2\rangle,|\tau_\uparrow,\sigma_\uparrow,\nu_1\rangle,|\tau_\uparrow,\sigma_\downarrow,\nu_4\rangle,|\tau_\downarrow,\sigma_\uparrow,\nu_4\rangle,|\tau_\downarrow,\sigma_\downarrow,\nu_1\rangle\},\\
					\text{At }C:&\quad \{|\tau_\uparrow,\sigma_\uparrow,\nu_1\rangle,|\tau_\uparrow,\sigma_\downarrow,\nu_4\rangle,|\tau_\downarrow,\sigma_\uparrow,\nu_4\rangle,|\tau_\downarrow,\sigma_\downarrow,\nu_1\rangle,|\tau_\uparrow,\sigma_\uparrow,\nu_2\rangle,|\tau_\uparrow,\sigma_\downarrow,\nu_3\rangle,|\tau_\downarrow,\sigma_\uparrow,\nu_3\rangle,|\tau_\downarrow,\sigma_\downarrow,\nu_2\rangle\},
				\end{align*}
			\end{widetext}
			then \eqref{bilayer Hamiltonian in Nambu} near $F$ and $G$ can be shown to have the form
			\begin{equation}\label{H0 F,G}
				H_0(\bm{F}_\pm)=v_F\delta k_x\tau_z+v_{\text{Ising}}\delta k_y\sigma_z\pm\Gamma\dfrac{1-\nu_z}{2}\tau_z.
			\end{equation}
			Similarly near $\bm{C}_\pm$ we have
			\begin{equation}\label{H0 C,D}
				H_0(\bm{C}_\pm)=v_F\delta k_x\tau_z+v_{\text{Ising}}\delta k_y\sigma_z\pm\Gamma\dfrac{1-\nu_z}{2}\sigma_z.
			\end{equation}
			Here we slightly abuse the notation: instead of introducing a new symbol, we use the $\nu$-Pauli matrices to represent the low ($\nu_z=1$) and high ($\nu_z=-1$) energy subspaces.
		\subsubsection{Low-energy Effective Theory in Nambu Space}
			The next step is to turn on an in-plane Zeeman exchange field and the intra-layer superconducting order parameters $\Delta$. With out loss of generality, below we consider the in-plane Zeeman field to be along the $k_y$ direction: $\bm{b}=(0,b_y)$, together with a \emph{spin-dependent hopping} $t_{s,\perp}=t_y$ due to a ferromagnetic layer:
			\begin{align}\label{H'}
				H'=b_y\sigma_y+\Delta\cos\frac{\varphi}{2}\cdot \sigma_y\tau_y+\Delta\sin\frac{\varphi}{2}\cdot\sigma_y\tau_x\nu_z+t_y\sigma_y\nu_x.
			\end{align}
			Here $\Delta$ is real and we add a Josephson phase difference $\varphi$ between the top and bottom layer through turning on the super-current.\par
			Still taking the basis we chose for the low-energy space for each node in the former section, we have
			\begin{widetext}
				\begin{align}\label{H' for Q+ and Q-}
					H'(\bm{F}_+)&=-b_y\sin\xi\cdot\sigma_y\nu_z+\Delta\cos\frac{\varphi}{2}\cdot\sigma_y\tau_y+\Delta\cos\xi\sin\frac{\varphi}{2}\cdot\sigma_x\tau_y\nu_z+b_y\cos\xi\cdot\sigma_y\nu_x+\Delta\sin\xi\sin\frac{\varphi}{2}\cdot\sigma_x\tau_y\nu_x+t_y\sigma_y,\\
					H'(\bm{F}_+)&=b_y\sin\xi\cdot\sigma_y\nu_z+\Delta\cos\frac{\varphi}{2}\cdot\sigma_y\tau_y-\Delta\cos\xi\sin\frac{\varphi}{2}\cdot\sigma_x\tau_y\nu_z+b_y\cos\xi\cdot\sigma_y\nu_x+\Delta\sin\xi\sin\frac{\varphi}{2}\cdot\sigma_x\tau_y\nu_x+t_y\sigma_y.
				\end{align}
				And for $\bm{C}_\pm$:
				\begin{align}
					H'(\bm{C}_+)&=b_y\cos\xi\cdot\sigma_y+\Delta\cos\frac{\varphi}{2}\cdot\sigma_y\tau_y+\Delta\cos\xi\sin\frac{\varphi}{2}\cdot\sigma_y\tau_x\nu_z-b_y\sin\xi\cdot\sigma_x\tau_z\nu_y+\Delta\sin\xi\sin\frac{\varphi}{2}\cdot\sigma_x\tau_y\nu_x\nonumber\\
					&\quad-\frac{\Delta}{2v_{\text{Ising}}}\sin^2\xi\sin\frac{\varphi}{2} \cdot v_2\theta\cdot\sigma_x\tau_y\nu_z+\frac{\Delta}{2v_{\text{Ising}}}\cos\xi\sin\xi \sin\frac{\varphi}{2} \cdot v_2\theta\cdot \sigma_y\tau_x\nu_x-\frac{t_y}{2v_{\text{Ising}}}\sin\xi\cdot v_2\theta\cdot\sigma_y\nu_z+t_y\sigma_y\nu_x,\label{H' for P+}\\
					H'(\bm{C}_-)&=b_y\cos\xi\cdot\sigma_y+\Delta\cos\frac{\varphi}{2}\cdot\sigma_y\tau_y-\Delta\cos\xi\sin\frac{\varphi}{2}\cdot\sigma_y\tau_x\nu_z+b_y\sin\xi\cdot\sigma_x\tau_z\nu_y+\Delta\sin\xi\sin\frac{\varphi}{2}\cdot\sigma_x\tau_y\nu_x \nonumber\\
					&\quad-\frac{\Delta}{2v_{\text{Ising}}}\sin^2\xi\sin\frac{\varphi}{2} \cdot v_2\theta\cdot\sigma_x\tau_y\nu_z -\frac{\Delta}{2v_{\text{Ising}}}\cos\xi\sin\xi \sin\frac{\varphi}{2} \cdot v_2\theta\cdot \sigma_y\tau_x\nu_x-\frac{t_y}{2v_{\text{Ising}}}\sin\xi\cdot v_2\theta\cdot\sigma_y\nu_z+t_y\sigma_y\nu_x.\label{H' for P-}
				\end{align}
				The effective Hamiltonian for the low-energy sector $|\nu_\uparrow\rangle$ then can be obtained from standard perturbation theory. We will keep the $\theta$-linear-order and the $\Delta$-second-order. Since both $\theta$ and $\Delta$ are small, we will drop terms proportional to $\Delta^2\theta$. Finally:
				\begin{align}
					H_{\text{eff}}(\bm{F}_\pm)&=v_F\delta k_x \tau_z+v_{\text{Ising}}\delta k_y \sigma_z\mp (b_y\sin\xi\mp t_y)\cdot\sigma_y+\Delta\cos\frac{\varphi}{2}\cdot\sigma_y\tau_y\pm\Delta\cos\xi\sin\frac{\varphi}{2}\cdot\sigma_x\tau_y\nonumber\\
					&\quad\pm\frac{1}{\Gamma}\Big[\big[-b_y^2\cos^2\xi+\Delta^2\sin^2\frac{\varphi}{2}\sin^2\xi\big]\tau_z+b_y\Delta\sin\frac{\varphi}{2}\sin 2\xi\sigma_z\tau_x\Big],\label{Heff for Q+ and Q-}\\
					H_{\text{eff}}(\bm{C}_\pm)&=v_F\delta k_x \tau_z+v_{\text{Ising}}\delta k_y \sigma_z+b_y\cos\xi\cdot\sigma_y+\Delta\cos\frac{\varphi}{2}\cdot\sigma_y\tau_y\pm\Delta\cos\xi\sin\frac{\varphi}{2}\cdot\sigma_y\tau_x\nonumber\\
					&\quad-\frac{\Delta v_2\theta}{2v_{\text{Ising}}}\sin^2\xi\sin\frac{\varphi}{2}\sigma_x\tau_y-\frac{t_y\cdot v_2\theta}{2v_{\text{Ising}}}\sin\xi\cdot\sigma_y\nonumber\\
					&\quad\pm\frac{1}{\Gamma}\Big[\big[b_y^2+\Delta^2\sin^2\xi\big]\sin^2\xi\cdot\sigma_z\pm2b_y\Delta\sin^2\xi\sin\frac{\varphi}{2}\cdot\sigma_z\tau_x+t_y^2\sigma_z\mp 2b_y t_y \sin\xi\cdot\tau_z\Big]\label{Heff for P+ and P-}.
				\end{align}
			\end{widetext}
		\subsubsection{Topological Nodes $C$ and $D$}
			Let us first focus on the node $C$ and $D$, i.e., $\bm{k}\sim\bm{C}_\pm$. It is convenient to perform a charge rotation $e^{\pm i\zeta/2\tau_z}$ to eliminate the $\pm\Delta\cos\xi\sin\frac{\varphi}{2}\cdot\sigma_y\tau_x$ term in \eqref{Heff for P+ and P-} with $\zeta\equiv\arctan(\cos\xi\tan\frac{\varphi}{2})$. In addition, the $\sigma_y$ term (and $\tau_z$ term) can be absorbed by redefining $\delta k_y$ (and $\delta k_x$). Using $e^{-i\zeta/2\tau_z}\tau_y e^{i\zeta/2\tau_z}=\cos\zeta\cdot\tau_y-\sin\zeta\cdot\tau_x$, we will arrive at
			\begin{widetext}
				\begin{align}
					\widetilde{H}_{\text{eff}}(\bm{C}_\pm)&=v_F\delta k_x \tau_z+v_{\text{Ising}}\delta k_y \sigma_z+\Big[b_y^{\text{origin.}}\cos\xi-\frac{t_y\cdot v_2\theta}{2v_{\text{Ising}}}\sin\xi\Big]\cdot\sigma_y+\tilde\Delta\cdot\sigma_y\tau_y\nonumber\\
					&\quad-\frac{\Delta v_2\theta}{2v_{\text{Ising}}}\sin^2\xi\sin\frac{\varphi}{2}\cdot\sigma_x\cdot(\cos\zeta\tau_y\mp\sin\zeta\tau_x)+\frac{2b_y\Delta}{\Gamma}\sin^2\xi\sin\frac{\varphi}{2}\cdot\sigma_z(\cos\zeta\cdot\tau_x\pm \sin\zeta\cdot\tau_y),\nonumber\\
					&=v_F\delta k_x \tau_z+v_{\text{Ising}}\delta k_y \sigma_z+b_y\cos\xi\cdot\sigma_y+\widetilde\Delta\cdot\sigma_y\tau_y\nonumber\\
					&\quad-\frac{\Delta v_2\theta}{2v_{\text{Ising}}}\sin^2\xi\sin\frac{\varphi}{2}\cdot\sigma_x\cdot(\cos\zeta\tau_y\mp\sin\zeta\tau_x)+\frac{2b_y\Delta}{\Gamma}\sin^2\xi\sin\frac{\varphi}{2}\cdot\sigma_z(\cos\zeta\cdot\tau_x\pm \sin\zeta\cdot\tau_y)\label{Heff_tilde for P+ and P-},
				\end{align}
			\end{widetext}
			with $\widetilde\Delta=\Delta\sqrt{\cos^2\frac{\varphi}{2}+\sin^2\frac{\varphi}{2}\cos^2\xi}\equiv\Delta\frac{\cos\frac{\varphi}{2}}{\cos\zeta}$. In \eqref{Heff_tilde for P+ and P-} we have denoted the original magnetic fields $b_y$ as $b_y^{\text{origin.}}$ and rewritten the $\sigma_y$ term as $b_y\cos\xi$ by introducing a new shifted field (this is why we introduce the spin-dependent tunneling together with the the magnetic fields)
			\begin{equation*}
				b_y\equiv b_y^{\text{origin.}}-\frac{t_y\cdot v_2\theta}{2v_{\text{Ising}}}\tan\xi=b_y^{\text{origin.}}-\frac{t_y\cdot v_2 t}{v_{\text{Ising}}^2 k_F}.
			\end{equation*}
			\indent Starting with the linear-order effective Hamiltonian (the first line of \eqref{Heff_tilde for P+ and P-}, so $\bm{C}_\pm$ has the same form), we find that a pair of Dirac nodes $\bm{k}^\pm_0=(\pm k^+_{x,0},0)$ along the $\delta k_x$ direction emerges when $b_y\cos\xi>\widetilde\Delta$. Repeating the two-band effective theory analysis (separating for $\sigma_y=\pm1$), we find, near the $\bm{k}_0^\pm$ Dirac nodes:
			\begin{equation}\label{heff for P+ and P-}
				h^\pm_{\text{eff}_\pm}(\bm{C})=\pm(-v_F\cos\theta^+\cdot\delta k_x\mu_z+v_{\text{Ising}}\sin\theta^+\cdot\delta k_y\mu_x),
			\end{equation}
			where
			\begin{equation*}
				\theta^+=\arctan\frac{\widetilde\Delta}{v_F k^+_{x,0}},\quad k^+_{x,0}=\frac{1}{v_F}\sqrt{(b_y\cos\xi)^2-\widetilde\Delta^2}
			\end{equation*}
			These effective theories are obtained in the basis:
			\begin{align*}
				|\psi^+_1\rangle&=|\sigma_y=1\rangle\otimes (i\sin\frac{\theta^+}{2}|\tau_z,\uparrow\rangle+\cos\frac{\theta^+}{2}|\tau_z,\downarrow\rangle),\\
				|\psi^+_2\rangle&=|\sigma_y=-1\rangle\otimes (\cos\frac{\theta^+}{2}|\tau_z,\uparrow\rangle-i\sin\frac{\theta^+}{2}|\tau_z,\downarrow\rangle),\\
				|\psi^-_1\rangle&=|\sigma_y=1\rangle\otimes(\cos\frac{\theta^+}{2}|\tau_z,\uparrow\rangle-i\sin\frac{\theta^+}{2}|\tau_z,\downarrow\rangle),\\
				|\psi^-_2\rangle&=|\sigma_y=-1\rangle\otimes (i\sin\frac{\theta^+}{2}|\tau_z,\uparrow\rangle+\cos\frac{\theta^+}{2}|\tau_z,\downarrow\rangle).
			\end{align*}
			To be concrete, states $|\sigma_y=\pm 1\rangle$ are defined as $|\sigma_y=~1\rangle=\frac{1}{\sqrt{2}}(|\sigma_z,\uparrow\rangle+i|\sigma_z,\downarrow\rangle)$ and $|\sigma_y=-1\rangle=\frac{1}{\sqrt{2}}(i|\sigma_z,\uparrow~\rangle+|\sigma_z,\downarrow\rangle)$.\par
			The left task to to project the last line of \eqref{Heff_tilde for P+ and P-} (as perturbation) onto these low-energy basis. We get, for $\bm{C}_+$, 
			\begin{align}
				\text{At }\bm{k}_0^+:\quad&\frac{\Delta v_2\theta}{2v_{\text{Ising}}}\sin^2\xi\sin\frac{\varphi}{2}(\sin\zeta\cos\theta^+\mu_x+\cos\zeta\mu_y)\nonumber\\
				&+\frac{2b_y\Delta}{\Gamma}\sin^2\xi\sin\frac{\varphi}{2}\cdot(-\cos\zeta\cos\theta^+\mu_y-\sin\zeta\mu_x),\label{H' for P+ at k0+}\\
				\text{At }\bm{k}_0^-:\quad&\frac{\Delta v_2\theta}{2v_{\text{Ising}}}\sin^2\xi\sin\frac{\varphi}{2}(\sin\zeta\cos\theta^+\mu_x-\cos\zeta\mu_y)\nonumber\\
				&+\frac{2b_y\Delta}{\Gamma}\sin^2\xi\sin\frac{\varphi}{2}\cdot(-\cos\zeta\cos\theta^+\mu_y+\sin\zeta\mu_x)\label{H' for P+ at k0-}
			\end{align}
			and for $\bm{C}_-$,
			\begin{align}
				\text{At }\bm{k}_0^+:\quad&\frac{\Delta v_2\theta}{2v_{\text{Ising}}}\sin^2\xi\sin\frac{\varphi}{2}(-\sin\zeta\cos\theta^+\mu_x+\cos\zeta\mu_y)\nonumber\\
				&+\frac{2b_y\Delta}{\Gamma}\sin^2\xi\sin\frac{\varphi}{2}\cdot(-\cos\zeta\cos\theta^+\mu_y+\sin\zeta\mu_x),\label{H' for P- at k0+}\\
				\text{At }\bm{k}_0^-:\quad&\frac{\Delta v_2\theta}{2v_{\text{Ising}}}\sin^2\xi\sin\frac{\varphi}{2}(-\sin\zeta\cos\theta^+\mu_x-\cos\zeta\mu_y)\nonumber\\
				&+\frac{2b_y\Delta}{\Gamma}\sin^2\xi\sin\frac{\varphi}{2}\cdot(-\cos\zeta\cos\theta^+\mu_y-\sin\zeta\mu_x).\label{H' for P- at k0-}
			\end{align}
			The new $\mu_x$-terms brought by perturbation just shift the nodes, while the new $\mu_y$-terms brought by perturbation open gaps. Collecting all $\mu_y$-terms, we get the four masses $m_\pm$ (each value is two-fold degenerate) for the nodes split from point intersection points $C$ and $D$:

			\begin{align}\label{topological mass gap}
				m_\pm(\bm{C}_\pm)&=\dfrac{2b_y\Delta}{\Gamma}\sin^2\xi\sin\frac{\varphi}{2}\cos\zeta\cos\theta^+\nonumber\\
				&\quad\times\left(1\pm\frac{c\theta\Gamma}{4b_yv_{\text{Ising}}\cos\theta^+}\right)\equiv m(1\pm\delta).
			\end{align}
			Clearly it is the dimensionless number $\delta\equiv\frac{c\theta\Gamma}{4b_yv_{\text{Ising}}\cos\theta^+}$ that dominates the topology around $\bm{C}_\pm$. If $|\delta|<1$, we find all four Dirac nodes generate masses of \emph{the same sign}, resulting in a total transfer of Chern Number $\pm4$. In terms of the original parameters, such condition reduces to
			\begin{equation}
				v_2t^2<\Delta k_Fv_{\text{Ising}}^2 a(\xi,\varphi)\sqrt{\left(\dfrac{b_y}{b_c^{\text{topo.}}}\right)^2-1},\label{eqn:v2_condition}
			\end{equation}
			with
			\begin{equation}\label{b_c topological}
				b_c^{\text{topo.}}\equiv\Delta\sqrt{\sec^2\xi\cos^2\frac{\varphi}{2}+\sin^2\frac{\varphi}{2}}.
			\end{equation}
			and
			\begin{equation*}
				a(\xi,\varphi)=\tan\xi\sin\xi\sqrt{\sec^2\xi\cos^2\frac{\varphi}{2}+\sin^2\frac{\varphi}{2}}.
			\end{equation*}
			As for the mass gap, in terms of original parameters, we have
			\begin{align}\label{topological mass gap (original paramter)}
				m&\equiv\dfrac{2b_y\Delta}{\Gamma}\sin^2\xi\sin\frac{\varphi}{2}\cos\zeta\cos\theta^+\nonumber\\
				&=\sin2\frac{\varphi}{2}\dfrac{\Delta^2\sin^2\xi}{v_{\text{Ising}} k_F\theta}\sqrt{\dfrac{(b_y/\Delta)^2}{\sec^2\xi\cos^2\frac{\varphi}{2}+\sin^2\frac{\varphi}{2}}-1}\nonumber\\
				&\equiv\sin\varphi\dfrac{\Delta^2\sin^2\xi}{v_{\text{Ising}} k_F\theta}\sqrt{\left(\dfrac{b_y}{b_c^{\text{topo.}}}\right)^2-1}.
			\end{align}
			Recall that $\varphi$ is the Josephson phase differences that we do no require to be small, so is $\xi=\arctan\frac{2t}{v_{\text{Ising}}k_F\theta}$. To have a meaningful topological gap, fraction $b_y/\Delta$ must be large enough so that the quantity inside the square root is positive definite. This requirement is satisfied for the two candidate $\mathrm{NbSe_2}$ and $\mathrm{TaS_2}$ we proposed in the main text --- which support large in-plane magnetic fields that far beyond the Pauli limit. 

		\subsubsection{Trivial Nodes $F$ and $G$}
			Finally let us look at the intersection point $F$ and $G$, i.e., $\bm{k}\sim\bm{F}_\pm$. Still we can absorb the last term of \eqref{Heff for Q+ and Q-} by redefining $\delta k_x$. After that we get
			\begin{align}\label{Heff_tilde for Q+ and Q-}
				\widetilde H_{\text{eff}}(\bm{F}_\pm)&=v_F\delta \tilde k_x \tau_z+v_{\text{Ising}}\delta k_y \sigma_z\mp (b_y^{\text{origin.}}\sin\xi\mp t_y)\cdot\sigma_y\nonumber\\
				&\quad+\Delta\cos\frac{\varphi}{2}\cdot\sigma_y\tau_y\pm\Delta\cos\xi\sin\frac{\varphi}{2}\cdot\sigma_x\tau_y\nonumber\\
				&\quad\quad\pm\frac{1}{\Gamma}b_y^{\text{origin.}}\Delta\sin\frac{\varphi}{2}\sin 2\xi\sigma_z\tau_x
			\end{align}
			Without bothering to perform the projection one more time, as we have done for $\bm{C}_\pm$, it's helpful to notice an effective ($\delta k_x\mapsto -\delta k_x$) mirror symmetry $\tau_y$ for the $\Delta$-linear order effective Hamiltonian $\widetilde H_{\text{eff}}(\bm{F}_\pm)$. Therefore the total Chern number must vanishes for the effective theory of each intersection point ($C$ and $D$). And the gap position can be easily solved from the first line of \eqref{Heff_tilde for Q+ and Q-}
			\begin{equation*}
				\tilde k_x=\pm\dfrac{1}{v_F}\sqrt{(b_y^{\text{origin.}}\sin\xi-t_y)^2-\Delta^2\cos^2\frac{\varphi}{2}},
			\end{equation*}
			with the gap size
			\begin{equation}\label{trivial mass gap}
				m(\bm{F}_\pm)=\Delta\cos\xi\sin\frac{\varphi}{2}.
			\end{equation}
			As a side remark, if one includes on the second-order perturbation, it can be shown that the second-order gap is always much smaller than the linear-order gap \eqref{trivial mass gap} in the current perturbative regime.

	\subsection{A different perturbative regime: $C=6$ Topological Phases}
	    In this section we briefly discuss the Chern number $C=6$ topological phases in a different perturbative regime. We start with the limit that the spin-independent hopping and the twisting angle both vanish: $t_\perp=0,\theta=0$. In addition, we let the spin-dependent hopping equal to the Zeeman exchange field: $t_y=b_y$ as in case-(i) in the main text.  
	    
	    Writing down the full linear order $k \cdot p$ BCS Hamiltonian near the point $P$:
	    \begin{align}
					H_{\text{BCS}}^{\text{bilayer}}(\bm{k})&=v_F k_x\tau_z+v_{\text{Ising}} k_y\sigma_z+\cos\frac{\varphi}{2} \Delta \sigma_y\tau_y+\sin\frac{\varphi}{2}\Delta\sigma_y\tau_x\nu_z\notag\\
					&-v_{\text{Ising}} k_F\theta/2 \sigma_z\nu_z+b_y\sigma_y+t_y\sigma_y\nu_x+t\tau_z\nu_x.\label{BCS Hamiltonian}
		\end{align}
		Precisely speaking, our strategy is to firstly consider the situation of $t_y=b_y\sim \Delta$ with $t_\perp=0,\theta=0$, and finally turn on a small $t_\perp\ll\Delta$ and a small twisting angle $v_{\text{Ising}}k_F\theta\ll \Delta$ as perturbations. This exactly corresponds to the regime in the numerical phase diagrams where $C=6$ is realized.
		
		One way to see the origin of the $C=6$ phase is to realize that when $t_y=b_y$, only the $\nu_x=1$ subspace (a half of the bands) experiences the magnetism captured by the two terms $b_y\sigma_y+t_y\sigma_y\nu_x$. Therefore, instead of 4 topological Dirac nodes (corresponding to the $C=12$ phase for the whole heterostructure), in the present situation Eqn.(\ref{BCS Hamiltonian}) only have 2 Dirac nodes emerging when $t_y=b_y$ is tuned up. After a small $t_\perp\ll\Delta$ and a small twisting angle $v_{\text{Ising}}k_F\theta\ll \Delta$ are turned on, these Dirac nodes receive topological mass gap and the $C=6$ phase is realized.

\section{Two-center approximation}
	Given two Bloch states $|\psi^\text{u}_{\bm{k},\alpha}\rangle$ and $|\psi^\text{d}_{\bm{k'},\beta}\rangle$ from top and bottom layer with the crystal momentum $\bm{k}$ and $\bm{k'}$ and sublattice labels $\alpha$ and $\beta$, two-center approximation \cite{bistritzer2011moire} tells that the tunneling strength between these two states takes a general form of \cite{catarina2019twisted}
	\begin{widetext}
		\begin{equation}\label{general tunneling}
			t^{\alpha\beta}_{\bm{k},\bm{k'}}\equiv\langle\psi^\text{u}_{\bm{k},\alpha}|H|\psi^\text{d}_{\bm{k'},\beta}\rangle=\dfrac{1}{V}\sum_{\bm{G}_1,\bm{G}_2}\delta_{\bm{k}+\bm{G}_1,\bm{k'}+\bm{G}_2}\cdot e^{-i\bm{G}_2\cdot\bm{\tau}_{2\beta}}\cdot t^{\alpha\beta}(\bm{k}+\bm{G}_1)\cdot e^{i\bm{G}_1\cdot\bm{\tau}_{1 \alpha}}
		\end{equation}
	\end{widetext}
	if one expands the Block state with Wannier basis for each layer, i.e., $t^{\alpha\beta}(\bm{k}-\bm{G}_2)\equiv\langle\bm{R}_1+\bm{\tau}_1|H|\bm{R}_2+\bm{\tau}_2\rangle$ with $\bm{\tau}_{1,2}$ the sublattice vectors, and makes use of the Poisson resummation formula \cite{ledwith2021lecture}.\par
	The allowed tunneling processes are constrained by the delta function in \eqref{general tunneling}, while its concrete form mainly comes from the two exponentials. More concretely, let us consider the bilayer system with the top layer rotated by a small angle $\theta/2$, and the bottom layer totated by $-\theta/2$, and focus on the region around the intersection point $P$: $\bm{k}=\bm{k}_P^{\text{t}}+\bm{F}^{\text{t}}$ and $\bm{k'}=\bm{k}_P^{\text{b}}+\bm{F}^{\text{b}}$ with $|\bm{F}^{\text{t,b}}|\ll1$. Since the real-space tunneling $t(\bm{r}^{\text{t}}-\bm{r}^{\text{b}})$ is the function of the spatial separation of two Wannier states $\sqrt{(\bm{r}^{\text{t}}-\bm{r}^{\text{b}})^2+|\bm{d}_\perp|^2}$, where inter-layer spacing $|\bm{d}_\perp|\gg\mathcal{O}(|\bm{r}^{\text{t}}-\bm{r}^{\text{b}}|)$, it should be flat enough in a large region of real space (to the same order as the Moir\'{e} pattern). Accordingly, $t^{\alpha\beta}(\bm{k}_P^{\text{t}}+\bm{F}^{\text{t}}+\bm{G}_1)\simeq t^{\alpha\beta}(\bm{k}_P^{\text{t}}+\bm{G}_1)$ is a good approximation and the summation over the BZ of each layer in \eqref{general tunneling} can be easily done:
	\begin{itemize}
		\item Since $|\bm{F}^\text{t}|, |\bm{F}^\text{b}|\gg1$, the Kronecker delta function in \eqref{general tunneling} is non-vanishing only if $\bm{G}_1$ and $\bm{G}_2$ differ by a small rotation. Namely,
			\begin{equation*}
				t^{\alpha\beta}_{\bm{F}^{\text{t}},\bm{F}^{\text{b}}}=\dfrac{1}{V}\sum_{\bm{G}_1,\bm{G}_2}\delta_{\bm{G}_1,\bm{G}_2}\cdot e^{-i\bm{G}_2\cdot\bm{\tau}_{2\beta}}\cdot t^{\alpha\beta}(\bm{k}_P+\bm{G}_1)\cdot e^{i\bm{G}_1\cdot\bm{\tau}_{1 \alpha}}
			\end{equation*}
		\item Since the three-fold intersection point $P$ are NOT connected with the reciprocal vectors for each layer (as a sharp contrast, in twisted bilayer graphene the three-fold $K$ valleys are directly connected with some reciprocal vectors --- this is actually how Moir\'{e} pattern enters in constraining the form of tunneling terms), the above summation only need to count the branch with $\bm{G}_1=\bm{G}_2=\bm{0}$.
	\end{itemize}
	As a result, we find that the tunneling strength only has a trivial momentum-dependence $t^{\alpha\beta}_{\bm{F}^{\text{t}},\bm{F}^{\text{b}}}=t$.\label{app:two-center approximation}
	
\bibliography{Ising_Top}
\bibliographystyle{apsrev} 
\end{document}